\RequirePackage{lineno}
\newif\ifpreprint%
\preprintfalse%
\ifpreprint%
	\documentclass[preprint,english,aps,prl,floatfix,amssymb,a4paper,10pt]{revtex4-2}
    
\else%
	\documentclass[twocolumn,english,aps,prl,floatfix,amssymb,a4paper,10pt]{revtex4-2}
\fi%
\usepackage{verbatim}
\usepackage{amsmath}
\usepackage{physics}
\usepackage{chemformula}
\usepackage{dsfont}
\usepackage{amssymb}
\usepackage{bbold}
\usepackage{graphicx}
\usepackage{braket}
\usepackage{bbold}
\usepackage{units}
\usepackage{lipsum}
\usepackage{xcolor}
\usepackage{hyperref}
\hypersetup{colorlinks = true, linkcolor=magenta,citecolor=blue, urlcolor=magenta, bookmarksnumbered =  true}

\renewcommand{\phi}{\varphi}

\renewcommand{\vec}[1]{\boldsymbol{#1}}

\newcommand{\crea}[1]{#1^{\dag}}
\newcommand{\anni}[1]{#1^{\vphantom{\dag}}}

\renewcommand{\braket}[2]{\langle#1|#2\rangle}
\newcommand{\ave}[1]{\langle#1\rangle}

\newcommand{\g}[1]{\textcolor{lightgray}{#1}}

\begin{document}

\ifpreprint%
	\linenumbers%
\fi%

\title{Stability of flat-band Bose-Einstein condensation from the geometry of compact localized states}

\author{
Kukka-Emilia Huhtinen}
\email{khuhtinen@phys.ethz.ch}
\affiliation{
Institute for Theoretical Physics, ETH Zurich, 8093 Zürich, Switzerland
}

\begin{abstract}
  We consider Bose-Einstein condensation in flat-band models from a
  real-space perspective. Using a basis of compact localized states,
  we reformulate the minimization of the mean-field energy as a
  Euclidian geometry problem. Within Bogoliubov theory, we show that flat-band
  models where the solutions to this problem are frameworks consisting
  of triangles with nonzero area are promising for condensation, whereas
  for instance square frameworks indicate condensation in a single
  mode is impossible. When restricting the analysis to Bloch states,
  this approach can be related to a necessary condition for a
  non-vanishing quantum distance. This work provides a new perspective
  on how condensation in flat bands is destabilized, and offers
  principles for the construction of models where flat-band
  Bose-Einstein condensation is possible.
\end{abstract}

\date{\today}

\maketitle

\section{Introduction}

Flat bands have attracted increased interest in recent years because
correlated phenomena are often enhanced in the absence of kinetic
energy~\cite{Torma2022,Zheng2014,Tasaki1998}. Naively, the infinite
effective mass of single particles might seem to present a problem for
phenomena such as superconductivity, but physical properties in
multiband models are affected by not only the band dispersion, but
also the geometry of quantum states. Of particular importance for
flat-band properties is the quantum metric~\cite{Provost1980} and its
basis-invariant generalizations, which can appear for instance in the
superfluid
weight~\cite{Peotta2015,Rossi2021,Torma2022,Peotta2023} and other electromagnetic
responses~\cite{Resta2011,Ahn2021,Topp2021,Verma2021,Hu2025,Tanaka2024,Komissarov2024,Verma2025}.

One interesting question is how the degeneracy of flat-band states
affects the properties of bosonic systems, and especially the
possibility of Bose-Einstein condensation. Bosonic flat-band systems
have been studied experimentally in various
platforms~\cite{Taie2015,Mukherjee2015,Vicencio2015,Kajiwara2016,Harder2020,Baboux2016,Whittaker2018,Scafirimuto2021,Harder2021,Jin2025,Wei2025,Li2025,Meng2023},
and previous theoretical research has indicated that some models such
as the kagome lattice can host a condensate for large enough
densities~\cite{Huber2010,You2012,Jalali-mola2023}, while others can
not. Quantum geometry gives a beginning of an answer regarding the
stability of flat-band condensates and superfluids: the quantum
distance and the quantum metric play an important role. The former is
involved in the excitation
fraction~\cite{Julku2021a,Julku2021b,Julku2023}, while the latter can
determine the speed of
sound~\cite{Julku2021a,Julku2021b,Julku2023,Lukin2023}, the effective
mass of superfluid carriers~\cite{Iskin2023} and part of the
superfluid weight~\cite{momentumspace}. A momentum-space approach
requires several assumptions, however: for instance, condensation is
typically assumed to occur in a Bloch state, which might not always be
the case in a flat band. To verify this assumption, a real-space
approach, such as the one used in Ref.~\onlinecite{You2012} for the
kagome lattice, is typically necessary.

Here, we adopt a related approach for generic tight-binding flat-band
models, relying on the existence of compact localized states
(CLSs). CLSs are single-particle flat-band eigenstates localized in
real space by destructive interference~\cite{Sutherland1986,Aoki1996,Rhim2021}, and give insight,
for instance, on band touchings between flat and dispersive
bands~\cite{Rhim2019,Bergman2008,Hwang2021}. CLSs have been observed
in various flat band
systems~\cite{Leykam2018a,Taie2015,Baboux2016,Tang2020,Mukherjee2015,Mukherjee2015b,Xia2016,Zong2016,Travkin2017,Lin2022,Shen2022,Hanafi2022,Song2023,Chase-Mayoral2024,Riva2025,Song2025},
and can be used as a starting point for the construction of flat band
models~\cite{Leykam2018a,Danieli2024,Danieli2026,Maimaiti2017,Maimaiti2019,Maimaiti2021,Maimaiti2021b,Hwang2021b,Graf2021,Kim2023,Liu2025b}.

In this work, we show how the properties of CLSs,
i.e. eigenstates of the non-interacting Hamiltonian, impact whether
bosons can condense once interactions are included. In this sense,
the method is similar in spirit to the usual quantum geometric
approach, which relates properties of an interacting system to the geometry of
Bloch states. However, the two approaches are not equivalent, but
rather complementary. The method proposed here considers all flat-band
states instead of only Bloch states, and because CLSs are a convenient
starting point in the construction of flat band models, offers
principles for designing flat bands with stable Bose-Einstein condensation.

\section{Bogoliubov theory on a flat band}

We consider the Bose-Hubbard Hamiltonian on a multiband lattice, $H = H_{\rm kin} + H_{\rm int}$, where
\begin{equation}
\begin{aligned}
H_{\rm kin} =& \sum_{i\alpha,j\beta} \crea{b_{i\alpha}} t_{i\alpha,j\beta} \anni{b_{j\beta}}, \\
H_{\rm int} =&  \frac{U}{2}\sum_{i\alpha}n_{i\alpha}(n_{i\alpha}-1)- \mu\sum_{i\alpha} n_{i\alpha}. \label{eq.bh-hamiltonian}
\end{aligned}
\end{equation}
Here, $\crea{b_{i\alpha}}$ creates a boson at orbital $\alpha$ in the
$i$th unit cell, and $n_{i\alpha} =
\crea{b_{i\alpha}}\anni{b_{i\alpha}}$. The bosons can hop from site
$j\beta$ to $i\alpha$ with amplitude $t_{i\alpha,j\beta}$, and
$t_{i\alpha,i\alpha}$ is the on-site potential at $i\alpha$. Particles
experience an on-site repulsive interaction $U$, and $\mu$ is the chemical potential. 

We assume condensation in a single flat-band eigenstate
$\ket{\varphi_0}$ such that $H_{\rm kin}\ket{\varphi_0} =
\varepsilon_0\ket{\varphi_0}$, where $\varepsilon_0$ is the flat-band
energy, and study the stability of the condensate within Bogoliubov
theory. The state $\ket{\varphi_0}$ does not need to be a Bloch
state. We expand $b_{i\alpha} =
\sqrt{N_0}\varphi_{i\alpha}+c_{i\alpha}$, where $\varphi_{i\alpha}\equiv
\ket{\varphi_0(\vec{r}_{i\alpha})}$ with $\vec{r}_{i\alpha}$ the
position of site $i\alpha$, and $N_0$ is the number of
condensed particles. The bosonic operator $\anni{c}_{i\alpha}$ annihilates
fluctuations on top of the condensate. By retaining only terms
quadratic in the fluctuations, we obtain the Hamiltonian
\begin{equation}
\begin{aligned}
    \frac{H}{N} =& \frac{1}{2N} \vec{\psi}^{\dag} H_B \vec{\psi} + E_c \\
    H_B =& \begin{pmatrix} \label{eq.ham}
    H_{\rm kin} - \mu_{\rm eff} & \Delta \\
    \Delta^{\dag} & H_{\rm kin}^* - \mu_{\rm eff}
    \end{pmatrix}, \\
    \vec{\psi}^{\dag} =& \left(\crea{c_{11}},\ldots \crea{c_{N_{\rm
          c},N_{\rm b}}}, \anni{c_{11}}, \ldots, \anni{c_{N_{\rm
          c},N_{\rm b}}} \right),
\end{aligned}
\end{equation}
where $[\Delta]_{i\alpha,j\beta} = UNn_0\varphi_{i\alpha}^2
\delta_{i\alpha,j\beta}$ and $[\mu_{\rm eff}]_{i\alpha,j\beta} =
\mu-2UNn_0|\phi_{i\alpha}|^2\delta_{i\alpha,j\beta}$. We define
$n_0=N_0/N$, where $N=N_cN_{\rm b}$ is the total number of sites,
$N_{\rm b}$ the number of bands and $N_{\rm c}$ the number unit
cells. The energy $E_c$ relates to the ground state energy of the
condensate~\cite{supplementary}.

The state $\ket{\varphi_0}$ is determined by minimizing the mean-field
energy
\begin{equation}
  E_{\rm MF} = n_0 \bra{\varphi_0}H_{\rm kin} \ket{\varphi_0} -
  n_0\mu + N\frac{Un_0^2}{2}\sum_{i\alpha}|\varphi_{i\alpha}|^4
\end{equation}
while enforcing the normalization $\braket{\varphi_0}{\varphi_0}=1$.  In a
flat band, $\bra{\varphi_0}H_{\rm kin}\ket{\varphi_0}=\varepsilon_0$
is a constant, and we simply need to minimize
$\sum_{i\alpha}|\varphi_{i\alpha}|^4$. Whenever possible,
$\ket{\varphi_0}$ will therefore have uniform densities,
$|\varphi_{i\alpha}|=1/\sqrt{N}$. This work focuses on models where at
least one such uniform-density state exists on the flat band.

The Hamiltonian~\eqref{eq.ham} can be written in a diagonal form with
$\crea{\vec{\psi}}H_B\anni{\vec{\psi}} = \crea{\vec{\gamma}}
\vec{\varepsilon} \anni{\vec{\gamma}}$, where $\crea{\vec{\gamma}} =
(\crea{\gamma_{1+}},\ldots,\crea{\gamma_{N+}},\anni{\gamma_{1-}},\ldots,\anni{\gamma_{N-}})$
contains Bogoliubov boson operators and $\vec{\varepsilon}$ the
corresponding energies. Because
$\crea{\gamma_{i\sigma}},\anni{\gamma_{i\sigma}}$ need to fulfill
bosonic commutation relations, we carry out a paraunitary
standardization of $H_B$~(see
Refs.~\cite{Colpa1978,Colpa1986a,Colpa1986b} and supplementary
material~\cite{supplementary}). In practice, $\crea{\vec{\gamma}}$ and
$\vec{\varepsilon}$ can be determined from the Jordan normal form of
the non-Hermitian matrix $\gamma_zH_B$, with $\gamma_z=\sigma_z\otimes
\mathbb{1}_N$~\cite{Xu2020}. In our case, $\gamma_zH_B$ has real
eigenvalues and a one-dimensional kernel spanned by
$(\ket{\varphi_0},-\ket{\varphi_0^*})^T$. Additional vanishing
eigenvalues would indicate that condensation is not possible in
$\ket{\varphi_0}$ alone, at least within traditional Bogoliubov
theory.

Keeping this important point in mind, let us assume that there exists
a pair of flat-band eigenstates $\ket{\varphi_0'} = C\ket{\varphi_0}$
and $\ket{\varphi_0''} = C^{\dag}\ket{\varphi_0}$, where $C$ is a
diagonal complex matrix, $[C]_{i\alpha,i\alpha}=a_{i\alpha}e^{i\xi_{i\alpha}}$, with
$a_{i\alpha}$ and $\xi_{i\alpha}$ real numbers. Then 
$\varphi_{i\alpha}'=e^{2i\xi_{i\alpha}}\varphi_{i\alpha}'' = a_{i\alpha}e^{i\xi_{i\alpha}}\varphi_{i\alpha}$. The key observation is that  
\begin{equation}
  \begin{aligned}
      &[(H_{\rm kin}-\mu_{\rm
          eff})\ket{\varphi_0'}-\Delta\ket{\varphi_0''^*}]_{i\alpha} \\
      =& \varepsilon_0\varphi_{i\alpha}'-\mu_{\rm
        eff}\varphi_{i\alpha}' -
      UNn_0\varphi_{i\alpha}^2(\varphi_{i\alpha}'')^* \\
      =& Un_0\left(1 - \frac{N|\varphi_{i\alpha}'|^2}{a_{i\alpha}^2}
      e^{-2i\xi_{i\alpha}}e^{2i\xi_{i\alpha}}
      \right)\varphi_{i\alpha}' = 0. \label{eq.zero_eigenvalue} 
  \end{aligned}
\end{equation}
On the second line, we have used that $\ket{\varphi_0'}$ is a flat-band
eigenstate, i.e. $H_{\rm kin}\ket{\varphi_0'}=\varepsilon_0\ket{\varphi_0'}$. On the third line, we have plugged in the effective chemical
potential for a uniform-density flat-band condensate, $\mu_{\rm
  eff}=\varepsilon_0-Un_0$~\cite{supplementary},
and used the fact that $|\varphi_{i\alpha}'|^2 = a_{i\alpha}^2/N$. A
similar reasoning shows that $[(H_{\rm kin}^*-\mu_{\rm
    eff})\ket{\varphi_0''^*} - \Delta^{\dag}\ket{\varphi_0'}]=0$. 

It follows that
$\gamma_zH_B(\ket{\varphi_0'},-\ket{\varphi_0''^*})^{T} = 0$: the
kernel of $\gamma_zH_B$ contains at least both
$(\ket{\varphi_0},-\ket{\varphi_0}^*)^T$ and
$(\ket{\varphi_0'},-\ket{\varphi_0''}^*)^T$. When $C$ is not
proportional to the identity matrix, $\ket{\varphi_0}$,
$\ket{\varphi_0'}$ and $\ket{\varphi_0''}$ are not simply related by a
gauge transformation, and the kernel of $\gamma_z H_B$ is at least
two-dimensional. This goes against the assumption of condensation in a
single mode.

On a flat band, for a stable condensate in $\ket{\varphi_0}$,
excluding gauge transformations, there should therefore exist no pairs
of flat-band states $C\ket{\varphi_0}$, $C^{\dag}\ket{\varphi_0}$ with
$C$ a diagonal matrix. In the particular case that there exists a
state $R\ket{\varphi_0}$, where $R$ is a real diagonal matrix not
proportional to the identity, this single state is sufficient to
destabilize condensation in $\ket{\varphi_0}$. It is important to note
that $C$ and $R$ do not need to be unitary: the states destabilizing
condensation do not need to minimize $E_{\rm MF}$, emphasizing that
flat-band condensation can be destabilized by geometric effects even
when $E_{\rm MF}$ has a non-degenerate minimum.

\begin{figure}
  \includegraphics[width=\columnwidth]{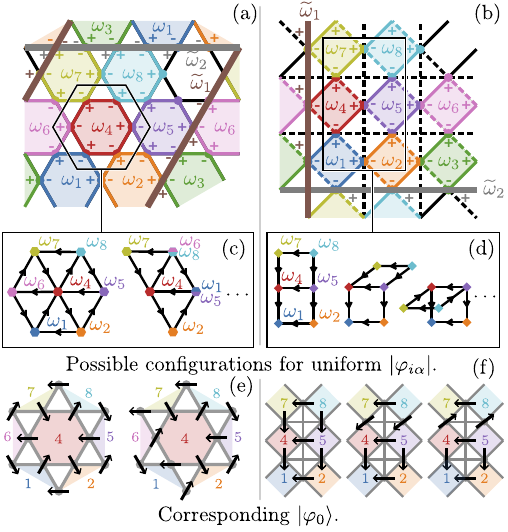}
  \caption{CLSs (hexagons/squares of different colors) and NLSs (thick
    brown/gray lines) in (a) the kagome
    lattice and (b) the checkerboard lattice, in a system of $3\times 3$
  unit cells with periodic boundary conditions. Full lines
  indicate a hopping amplitude $t=1$, while dashed lines indicate
  $t=-1$. Considering the 
  overlaps of CLSs at sites within the region bounded by a thin black
  line, and requiring uniform $|\varphi_{i\alpha}|$, we obtain a set of
  constraints $|\omega_i-\omega_j| = 1/\sqrt{N}$ for
  neighboring $\omega_i$ and $\omega_j$, fixing the distance between
  these points in the complex plane. (c-d) Examples of possible
  configurations of coefficients $\omega_i$ in the complex plane in
  the (c) kagome and (d) checkerboard lattices. Each edge of the frameworks
  has length $1/\sqrt{N}$, and corresponds to $\varphi_{i\alpha}$ at a
  given site. The resulting $\ket{\varphi_0}$ are shown in
  (e-f).}
  \label{fig.cls_stability}
\end{figure}

\section{Compact localized states and flat-band condensation}

To determine whether condensation is possible on a flat band, we first
need to find the possible states $\ket{\varphi_0}$, and then to check
whether other problematic states $C\ket{\varphi_0}$,
$C^{\dag}\ket{\varphi_0}$ exist for each of these states. To this end,
we expand the flat-band states in terms of CLSs and non-contractible
loop states (NLS). CLSs are single-particle flat-band eigenstates
localized by destructive interference that have a nonzero amplitude in
a finite region of space, and vanish elsewhere (see
Fig.~\ref{fig.cls_stability}a-b). Because of the translational
symmetry of the lattice, we can associate a CLS $\ket{\nu_{i,n}}$ with
each unit cell $i$:
\begin{equation}
  \ket{\nu_{i,n}} =
  \sum_{j\alpha}\nu_{\vec{R}_j\alpha,n}\ket{(\vec{R}_{i}+\vec{R}_{j})\alpha}, 
\end{equation}
where $j$ runs over unit cells, $\vec{R}_i$ is the center of the $i$th
unit cell, and $\nu_{\vec{R}_j\alpha,n}$ are the amplitudes of the
CLSs associated with $\vec{R}_i=0$ at site $j\alpha$. The index $n$
runs over degenerate flat bands, $n=1\ldots N_{\rm fb}$, i.e. for
$N_{\rm fb}$ degenerate bands, there are $N_{\rm fb}$ CLSs associated
with each unit cell. CLSs can be constructed for any flat band of a
tight-binding model with finite-range hopping
processes~\cite{Rhim2019}.

While we can associate one CLS to each unit cell, they are not always
linearly independent in systems with periodic boundary
conditions~\cite{Bergman2008,Rhim2019}. This occurs in so-called
singular bands~\cite{Rhim2019,Rhim2021}, which can not be spanned by
CLSs alone and require NLSs (see Fig.~\ref{fig.cls_stability}a-b) or
other extended states. We refer to these additional states as
$\ket{\widetilde{\nu}_a} = \sum_{i\alpha}
\widetilde{\nu}_{i\alpha,a}\ket{\vec{R}_i\alpha}$, where $a$ labels the
states. The states $\ket{\nu_{i,n}}$ and $\ket{\widetilde{\nu}_a}$ fully
span the flat bands, and all eigenstates with $H_{\rm kin}\ket{\psi} =
\varepsilon_0\ket{\psi}$ can be written as a linear combination
\begin{equation}
  \ket{\psi} = \sum_{i,n}\omega_{i,n}\ket{\nu_{i,n}} + \sum_a
  \widetilde{\omega}_a\ket{\widetilde{\nu}_a}. \label{eq.wf_expansion}
\end{equation}
The choice of CLSs and NLSs is not unique, but any complete basis is
valid for the following arguments~\cite{supplementary}. 

As stated previously, we focus on systems where
there exists at least one flat-band eigenstate with uniform
$|\varphi_{i\alpha}| = 1/\sqrt{N}$, which is guaranteed to minimize
$E_{\rm MF}$. All such states fulfill the constraints
\begin{equation}
  \left| \sum_{j,n} \nu_{(\vec{R}_i-\vec{R}_j)\alpha,n}\omega_{j,n} +
  \sum_a\widetilde{\nu}_{i\alpha,a}\widetilde{\omega}_a  \right| =
  \frac{1}{\sqrt{N}}, \label{eq.constraint} 
\end{equation}
obtained by requiring that for each $i\alpha$, $|\psi_{i\alpha}|=1/\sqrt{N}$ in
Eq.~\eqref{eq.wf_expansion}. The
problem of finding all states minimizing $E_{\rm MF}$ is equivalent to the
geometric problem of finding the possible positions of $\omega_{j,n}$ and
$\widetilde{\omega}_a$ in the complex plane while fulfilling all the
constraints~\eqref{eq.constraint}.

As two simple examples, let us consider the kagome and checkerboard
lattices (Fig.~\ref{fig.cls_stability}). These lattices both feature a
lowest flat band with a singular band
touching~\cite{supplementary}. We first consider
sites where two CLSs overlap with opposite signs and NLSs are not
involved (Fig.~\ref{fig.cls_stability}a-b). The constraints~\eqref{eq.constraint} corresponding to these
sites acquire the form $|\omega_i-\omega_j|=1/\sqrt{N}$, where we have
ommited the index $n$ since we consider a single flat band. The distance
in the complex plane between $\omega_i$ and $\omega_j$ corresponding
to neighboring CLSs is therefore $1/\sqrt{N}$. In the kagome lattice,
this is fulfilled only when the coefficients $\omega_i$ are the
vertices of a triangular framework (see
Fig.~\ref{fig.cls_stability}c), while on the checkerboard lattice, the
coefficients lie on a square framework (see
Fig.~\ref{fig.cls_stability}d). Each edge of these frameworks
corresponds to $\varphi_{i\alpha}$ at one site (see
Fig.~\ref{fig.cls_stability}e-f). When also sites involving NLSs are
taken into account, the configurations shown in
Fig.~\ref{fig.cls_stability}(c-d) become compatible with periodic
boundary conditions~\cite{supplementary}.

Because only the distances between neighboring coefficients are fixed,
there are many configurations of $\omega_i$ fulfilling all
constraints~\eqref{eq.constraint} in both lattices. Some operations
have no physical impact: translating the whole framework does not
change the directions of any of the edges, and thus does not change
the corresponding wavefunction, while an overall rotation is a
gauge transformation. However, other transformations give a new
uniform-density flat-band eigenstate.  In the kagome lattice, while
the triangles can not be deformed without changing edge lengths (or
equivalently, without changing $|\varphi_{i\alpha}|$), two triangles
sharing an edge can point in opposite or identical directions. In
other words, "folding" or
"unfolding" along edges of the triangular framework yields new
configurations minimizing $E_{\rm MF}$ (see
Fig.~\ref{fig.cls_stability}(c)). In the checkerboard lattice, angles
between edges of the framework can be changed continuously without
affecting edge lengths (see Fig.~\ref{fig.cls_stability}(d)),
reflecting that states minimizing $E_{\rm MF}$ are connected by
continuous transformations which leave $E_{\rm MF}$ unchanged.

In the kagome and checkerboard lattices, Eq.~\eqref{eq.constraint}
simply constrains distances between coefficients $\omega_i$. In more
complicated cases, Eq.~\eqref{eq.constraint} constrains the distance
between linear combinations of several $\omega_i$. Flat band states
can then be represented as frameworks where each edge corresponds to
$b_{i\alpha}\varphi_{i\alpha}$, with $b_{i\alpha}$ a scalar, and each
vertex is a linear combination of coefficients $\omega_i$.

This approach allows us to geometrically determine all flat-band
states which minimize $E_{\rm MF}$. To verify that condensation is
possible, we now need to ensure for these $\ket{\varphi_0}$ that there
exist no problematic states $C\ket{\varphi_0}$,
$C^{\dag}\ket{\varphi_0}$.  Since all flat-band states are spanned by
CLSs and NLSs, they can be represented by frameworks with the same
underlying graph, but potentially different edge lengths, as the ones
representing the states minimizing $E_{\rm MF}$. Every complex
diagonal matrix can be rewritten as $C=RU$, where $R$ is a real
diagonal matrix and $U$ is a unitary diagonal matrix. Multiplication
by a real diagonal matrix $R$ corresponds to changing the lengths of
edges while keeping them parallel to their initial
direction. Multiplication by a diagonal unitary matrix $U$ corresponds
to rotating each edge by an angle $\xi_{i\alpha}$. Condensation is
unstable in a given $\ket{\varphi_0}$ if, excluding overall 
rotations and uniform scaling, it is possible to construct 1) a
framework with all edges parallel to those of the framework
representing $\ket{\varphi_0}$ or 2) two frameworks with edges rotated
by $+\xi_{i\alpha}$, $-\xi_{i\alpha}$ respectively compared to their
original orientation, and both having identical edge lengths.

In the checkerboard lattice, condensation is impossible in any
uniform-density $\ket{\varphi_0}$: we can flip the direction of a
subset of edges without changing the orientation of others, yielding a
different framework with all edges parallel to their original
direction (see Fig.~\ref{fig.cls_stability}(d)). In the kagome
lattice, each $\ket{\varphi_0}$ is represented by a framework
consisting of only triangles with nonzero area, which we refer to as a
triangulated framework. For such frameworks, the problematic
frameworks mentioned above can not be constructed, and condensation
can occur~\cite{supplementary}. Other lattices where all possible
$\ket{\varphi_0}$ correspond to triangulated frameworks include for
instance the sawtooth ladder~\cite{supplementary}.

In complicated models, it might be difficult to directly visualize all
$\ket{\varphi_0}$ as frameworks. As outline in the supplementary
material, it is also possible to directly verify from a basis of CLSs
and NLSs whether states $C\ket{\varphi_0}$, $C^{\dag}\ket{\varphi_0}$
exist for a given $\ket{\varphi_0}$, without explicitly considering
frameworks~\cite{supplementary}. However, this requires the prior
determination of $\ket{\varphi_0}$. 

\begin{figure}
  \includegraphics[width=\columnwidth]{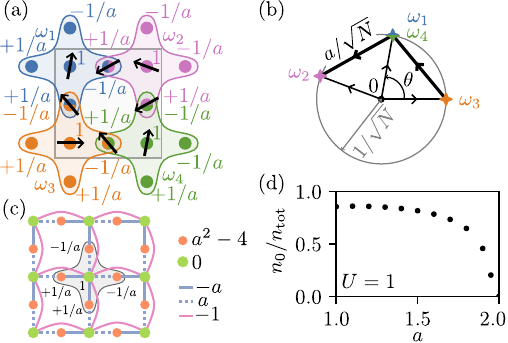}
  \caption{(a) CLSs used as a starting point for the construction of a
    model with stable flat-band condensation. Each CLS is centered on
    a site where there are no overlaps, which constrains
    $|\omega_i|=1/\sqrt{N}$ to achieve uniform densities. The
    amplitude of the CLS at other sites is $\pm 1/a$, and the overlaps
    with adjacent CLSs enforce
    $|\omega_i-\omega_j|=a/\sqrt{N}$. Together, these constraints then
    require that $0$, $\omega_i$ and $\omega_j$ form an isosceles triangle. (b)
    Possible configuration for the coefficients $\omega_i$, under the
    constraints related to the sites in the gray square in (a). The
    thin black edges have a length of $1/\sqrt{N}$, and correspond to
    $\varphi_{i\alpha}$ at the central site of a CLS. The thick black
    lines have a length of $a/\sqrt{N}$, and give $a\varphi_{i\alpha}$
    for the other sites. The $\varphi_{i\alpha}$ corresponding to this
    configuration are indicated by arrows in (a). (c) Tight-binding
    model for the Tasaki lattice considered here. (d) Condensate
    fraction $n_0/n_{\rm tot}$ assuming condensation in the
    uniform-density state with the lowest ZPE. We set $n_{\rm
      tot}=1$.} \label{fig.tasaki_model}
\end{figure}

\section{Tasaki lattice with stable flat-band condensation}

Based on our observations, we can construct CLSs that are compatible
 with condensation: we simply need to ensure that they overlap
 in such a way that uniform-density flat-band eigenstates are
 represented by triangulated frameworks. An example, as well as a
 possible configuration of $\omega_i$ that minimizes $E_{\rm MF}$, are
 shown in Fig.~\ref{fig.tasaki_model}a-b. We work on a Tasaki
 lattice~\cite{Tasaki1992}, with the tight-binding parameters shown in
 Fig.~\ref{fig.tasaki_model}c (note that tight-binding parameters for
 a given CLS are not unique). We introduce a tuning parameter $a$
 which controls the possible phases of $\omega_i$ when $E_{\rm MF}$ is
 minimized: in adjacent unit cells $i$ and $j$,
 $\omega_i=\omega_je^{\pm i\theta}$, where $\theta=
 2\arcsin(a/2)$. This implies that the Bloch states at
 $\vec{k}=2\arcsin(a/2)(\sigma,\sigma')$, with $\sigma,\sigma'=\pm
 1$ have uniform densities.

When using periodic boundary conditions, uniform-density flat-band states are
commensurate with the system size only for a few values of $a$ at
numerically accessible system sizes. We therefore use open boundary
conditions. In order to retain the perfect flatness of the band and
the uniformity of $\ket{\varphi_0}$, the tight-binding parameters at
the boundaries are adjusted (detailed parameters are given in the
supplementary material~\cite{supplementary}).

To determine which state is the most favorable for condensation, we
compute the zero-point energy (ZPE) of the Bogoliubov phonons
corresponding to randomly sampled uniform-density $\ket{\varphi_0}$. The
sampling is performed by taking advantage of a correspondence between
the possible $\ket{\varphi_0}$ and three-colorings of a square
lattice~\cite{supplementary}. The three-coloring sampling was
performed using the Wang-Swendsen-Koteck\'{y} (WSK)
algorithm~\cite{Wang1989,Wang1990}. Although this algorithm is not
always ergodic~\cite{Lubin1993,Mohar2009,Mohar2010}, the ergodicity is
guaranteed on bipartite lattices~\cite{Salas2022}, including the
square lattice.

The state $\ket{\varphi_0}$ with the lowest ZPE, which is the most
favorable for condensation, is a superposition of the $\vec{k}=(0,0)$
and $\vec{k}=(\pi,\pi)$ Bloch states, while the least favorable
$\ket{\varphi_0}$ are the Bloch states at
$\vec{k}_c=2\arcsin(a/2)(\sigma,\sigma')$ (see supplementary
material~\cite{supplementary} for details). Assuming condensation in
the most favorable state, $n_0$ is nonzero, signaling condensation is
possible, except in the limit $a\to 2$ (see
Fig.~\ref{fig.tasaki_model}d).

At $a=2$, the area of the triangles in Fig.~\ref{fig.tasaki_model}b
vanishes, and $E_{\rm MF}$ is minimized by only the $(\pi,\pi)$ Bloch
state. However, even though there is no degeneracy at the mean-field
level, condensation becomes unstable because of the existence of
problematic non-uniform flat-band
eigenstates~\cite{supplementary}. This highlights that also states
that do not minimize $E_{\rm MF}$ can destabilize the condensate due
to quantum geometric effects.

\section{Relationship to the condensate quantum distance}

Our method has some similarity to previous quantum geometric
approaches, since it predicts the possibility of a stable condensate
from the properties of the non-interacting model. Quantum geometry and
the properties of CLSs and NLSs can give complementary information for
instance in the context of singular bands: the necessity of NLSs
relates directly to the singularity of the band touching and the
existence of boundary states in systems with open boundary conditions,
while the quantum distance gives information on the strength of the
singularity~\cite{Rhim2020,Oh2022,Filusch2023}. Similarly, our method
complements previous research relating the quantum distance to
the excitation fraction.

When condensation is assumed to occur in a Bloch state $\ket{n_{\vec{k_c}}}$
corresponding to a momentum $\vec{k_c}$ on a single non-degenerate
flat band, it was shown in Refs.~\cite{Julku2021a,Julku2023} that the
excitation fraction is related to the so-called condensate quantum
distance $D(\vec{q})$ by
\begin{equation}
  \begin{aligned}
    \lim_{U\to 0}n_{\rm ex} &= \frac{1}{N_c}\sum_{\vec{q}\neq
      0} \frac{1-D(\vec{q})}{2D(\vec{q})}, \\
    D(\vec{q})&= \sqrt{1-|\delta(\vec{q})|^2}, \\
    \delta(\vec{q}) &=
    N_b\sum_{\alpha}\braket{n_{\vec{k_c}+\vec{q}}}{\alpha}\braket{n_{\vec{k_c}-\vec{q}}}{\alpha}\braket{\alpha}{n_{\vec{k_c}}}^2, 
  \end{aligned}
\end{equation}
where $\ket{n_{\vec{k}}}$ is the periodic part of the Bloch function
on the flat band and
$|\braket{\alpha}{n_{\vec{k_c}}}|^2=1/N_b$, i.e. $\ket{\varphi_0}$ is
again assumed to have uniform densities. In real space,
$\ket{n_{\vec{k}}}$ corresponds to a flat-band eigenstate such that
$\varphi_{i\alpha} =
\braket{\alpha}{n_{\vec{k}}}e^{i\vec{k}\cdot\vec{r}_{i\alpha}}/\sqrt{N_c}$.
The condensate quantum distance reduces to the usual quantum distance
$d(\vec{q}) =
\sqrt{1-|\braket{n_{\vec{k_c}+\vec{q}}}{n_{\vec{k_c}-\vec{q}}}|^2}$
when all Bloch states are real.

In this work, we determine whether states $\ket{\varphi_0'}=C\ket{\varphi_0}$,
$\ket{\varphi_0''}=C^{\dag}\ket{\varphi_0}$ exist on the flat
band for any given $\ket{\varphi_0}$. When $\ket{\varphi_0}$ is a Bloch state,
$\varphi_{i\alpha}'=a_{i\alpha}e^{i\vec{k_c}\cdot
  \vec{r}_{i\alpha}}e^{i\xi_{i\alpha}}/\sqrt{N}$, $\varphi_{i\alpha}''=a_{i\alpha}e^{i\vec{k_c}\cdot
  \vec{r}_{i\alpha}}e^{-i\xi_{i\alpha}}/\sqrt{N}$. These two states
will also be Bloch states when $a_{i\alpha}=a_{\alpha}$ is
independent of the unit cell, and $\xi_{i\alpha}=\xi_{\alpha}+\vec{q}\cdot
\vec{r}_{i\alpha}$ for some momentum $\vec{q}$. In other words, if we
require that  $\ket{\varphi_0}$, $\ket{\varphi_0'}$ and
$\ket{\varphi_0''}$ are Bloch states, $\ket{\varphi_0'}$ and
$\ket{\varphi_0''}$ will correspond to Bloch states
$\ket{n_{\vec{k_c}+\vec{q}}}$ and $\ket{n_{\vec{k_c}-\vec{q}}}$ with periodic parts
$\ket{n_{\vec{k_c}+\vec{q}}} =
\tilde{C}\ket{n_{\vec{k_c}}}$, $\ket{n_{\vec{k_c}-\vec{q}}} =
\tilde{C}^{\dag}\ket{n_{\vec{k_c}}}$, where $\tilde{C}$ is a diagonal
matrix. If such states indeed exist,
\begin{equation}
  \begin{aligned}
    \delta(\vec{q}) =
    \frac{1}{N_b}\sum_{\alpha}\tilde{C}_{\alpha\alpha}^2 = 1,
  \end{aligned}
\end{equation}
where we have used that
$\braket{n_{\vec{k_c}+\vec{q}}}{n_{\vec{k_c}+\vec{q}}} =
(1/N_b)\sum_{\alpha}\tilde{C}_{\alpha\alpha}^2 =1$. It follows that
the condensate quantum distance vanishes at $\vec{q}$, and $n_{\rm
  ex}$ diverges. When restricting the analysis to Bloch states, the
real-space approach used here thus relies on a necessary condition for
the quantum distance to be nonzero at all $\vec{q}\neq 0$. This
approach is however more general, since it also accounts for states
that might destabilize the condensate but are not Bloch states.

\section{Discussion}

We have shown how the possibility of stable
condensation on a flat band is influenced by how CLSs overlap, which
determines constraints for states minimizing the mean-field
energy. These states can be visualized as frameworks, and a
triangulated framework is promising for condensation. Based on this
observation, it is possible to construct flat-band models with
potential for a stable Bose-Einstein condensate by using an
advantageous basis of CLSs as a starting point. This
method could be used for instance to build models that can be
implemented experimentally, or that have desired properties such as
topologically nontrivial flat bands.

This work provides further
insight into bosonic flat-band systems, complementing previous
research relating the stability of flat-band condensates and
superfluids to quantum geometric
quantities~\cite{Julku2021a,Julku2021b,Julku2023,Lukin2023,Iskin2023}. It
re-emphasizes the important impact of the geometry of eigenstates,
sometimes including those without translational invariance, on
flat-band properties.

This method could also give insight into phases occuring at
temperatures where Bose-Einstein condensation does not yet
occur. Previously, a trion phase has been predicted above the
temperature for Bose-Einstein condensation in the kagome
lattice~\cite{You2012}. This is related to the structure of the
flat-band eigenstates minimizing the mean-field energy: the phases of
uniform-density state are constrained to multiples of $2\pi/3$. In
this work, we constructed an example of a flat band model where the
possible phases are still constrained, but can take more than three
distinct values. Similar constraints on possible phases are likely to
occur in many flat bands with a stable condensate, since stability has
a relation to triangulated frameworks, where edge direction can not be
continuously modified. Similarly to the kagome lattice, such
constraints could manifest for instance in time-of-flight
measurements~\cite{You2012,Donini2024}.

\begin{acknowledgments}
  {\it Acknowledgments ---} The author thanks Sebastian D. Huber and
  Friederike Bartels for fruitful discussion, and Matteo Dürrnagel for
  helpful comments. The author gratefully acknowledges support by an ETH
  Zürich Postdoctoral Fellowship.
\end{acknowledgments}

\clearpage

\pagebreak
\setcounter{equation}{0}
\setcounter{figure}{0}
\setcounter{table}{0}
\setcounter{section}{0}
\setcounter{page}{1}
\makeatletter
\renewcommand{\theequation}{S\arabic{equation}}
\renewcommand{\thefigure}{S\arabic{figure}}

\begin{widetext}

\begin{center}
  {\bf Supplemental material: Stability of flat-band Bose-Einstein condensation from the geometry of compact localized states} \\
  \vspace{0.5cm}
  Kukka-Emilia Huhtinen \\
  {\it 
    Institute for Theoretical Physics, ETH Zurich, 8093 Zürich, Switzerland
  }
\end{center}
\end{widetext}

\section{Band structures of considered lattice models}

\begin{figure}
  \includegraphics[width=\columnwidth]{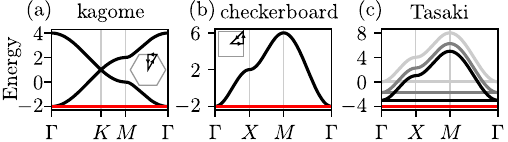}
  \caption{Band structures of the (a) kagome lattice, (b) checkerboard
  lattice and (c) modified Tasaki lattice. In the Tasaki lattice, we
  show band structures for $a=1$ (black), $a=1.5$ (gray) and $a=2$
  (light gray). In all panels, the flat band of interest is plotted in
  red.}
  \label{fig.band_structures}
\end{figure}

The band structure is obtained by Fourier transforming $H_{\rm
  kin}$
in a system with periodic boundary conditions and diagonalizing the
resulting matrix $[H_{\vec{k}}]_{\alpha\beta}=\sum_{j} t_{0\alpha,j\beta}e^{i\vec{k}\cdot
  \vec{R}_j}$, where $t_{i\alpha,j\beta}$ is the hopping amplitude between sites
$i\alpha$ and $j\beta$, and $\vec{R}_j$ is the center of the $j$:th
unit cell. In the kagome lattice, the Fourier 
transformation is given by
\begin{equation}
  H_{\vec{k}}^{\rm kagome} = \begin{pmatrix}
    0 & 1 + e^{i k_1} & 1 + e^{i k_3} \\
    1 + e^{-ik_y} & 0 & 1 + e^{-i k_2} \\
    1+ e^{-ik_3} & 1+e^{ik_2} & 0
  \end{pmatrix},
\end{equation}
where $k_1=k_x$, $k_2 = k_x/2+\sqrt{3}k_y/2$, $k_3=k_x/2-\sqrt{3}k_y/2$,
giving the band structure shown in
Fig.~\ref{fig.band_structures}(a). This model features a lowest flat
band with a singular band touching at $\Gamma$.

The Fourier-transformed kinetic Hamiltonian for the checkerboard model
considered in the main text is
\begin{equation}
  \begin{aligned}
    &H_{\vec{k}}^{\rm checkerboard} = \\
    &\begin{pmatrix}
    -2\cos(k_x) & 1+e^{-i(k_x-k_y)} - e^{ik_y} - e^{-ik_x}\\
    1+e^{i(k_x-k_y)} - e^{-ik_y} - e^{ik_x} & -2\cos(k_y) 
  \end{pmatrix}.
  \end{aligned}
\end{equation}
Like the kagome lattice, this lattice features a lowest flat band with
a singular band touching at $\Gamma$ (see
Fig.~\ref{fig.band_structures}(b)), but in contrast to the kagome
lattice, the flat band can not host a stable condensate. 

For the modified Tasaki lattice,
\begin{equation}
  H_{\vec{k}}^{\rm Tasaki} = \begin{pmatrix}
    -2(\cos(k_x)+\cos(k_y)) & a(1-e^{ik_x}) & a(1-e^{ik_y}) \\
    a(1-e^{-ik_x})& a^2-4 & 0 \\
    a(1-e^{-ik_y})& 0 & a^2-4
  \end{pmatrix}.
\end{equation}
In this model, for all $a$, there is a lowest isolated flat band at
energy $-4$, and a flat and dispersive band above the band gap. The
band stucture is plotted for $a=1$, $a=1.5$ and $a=2$ in
Fig.~\ref{fig.band_structures}(c). As $a$ is increased, the flat and
dispersive bands above the flat band of interest shift upwards in
energy, but there are no drastic changes in dispersions
otherwise. However, condensation rapidly destabilizes when
approaching $a=2$ due to quantum geometric effects.

\section{Details on Bogoliubov theory}

As stated in the main text, we consider the Bose-Hubbard model
$H=H_{\rm kin}+H_{\rm int}$, where
\begin{equation}
\begin{aligned}
H_{\rm kin} =& \sum_{i\alpha,j\beta} \crea{b_{i\alpha}} t_{i\alpha,j\beta} \anni{b_{j\beta}}, \\
H_{\rm int} =&  \frac{U}{2}\sum_{i\alpha}n_{i\alpha}(n_{i\alpha}-1)- \mu\sum_{i\alpha} n_{i\alpha}.
\end{aligned}
\end{equation}
As in the main text, $\anni{b_{i\alpha}}$ creates a boson at site
$i\alpha$, $t_{i\alpha,j\beta}$ is the hopping amplitude from site
$j\beta$ to $i\alpha$, $\mu$ is the chemical potential and $U$ the
on-site repulsive interaction. We consider condensations in a single
flat-band eigenstate $\ket{\varphi_0}$ with $H_{\rm
  kin}\ket{\varphi_0} = \varepsilon_0\ket{\varphi_0}$. 
By expanding $b_{i\alpha} = \sqrt{N_0}\varphi_{i\alpha}+c_{i\alpha}$
and retaining only quadratic terms in the fluctuations $c_{i\alpha}$,
we obtain
\begin{equation}
\begin{aligned}
    \frac{H}{N} =& \frac{1}{2N} \vec{\psi}^{\dag} H_B \vec{\psi} +
    E_{\rm MF} + E_0 \\ 
    H_B =& \begin{pmatrix}
    H_{\rm kin} - \mu_{\rm eff} & \Delta \\
    \Delta^{\dag} & H_{\rm kin}^* - \mu_{\rm eff}
    \end{pmatrix}, \\
    \vec{\psi}^{\dag} =& \left(\crea{c_{11}},\ldots \crea{c_{N_{\rm
          c}N_{\rm b}}}, \anni{c_{11}}, \ldots, \anni{c_{N_{\rm
          c}N_{\rm b}}} \right) \\
    E_{\rm MF} =& n_0 \bra{\varphi_0}H_{\rm kin} \ket{\varphi_0} - n_0\mu + N\frac{Un_0^2}{2}\sum_{i\alpha}|\varphi_{i\alpha}|^4, \\
    E_0 =& -\frac{1}{2}{\rm Tr}[H_{\rm kin}] + \frac{1}{2}\mu - Un_0,
\end{aligned}
\end{equation}
where $[\Delta]_{i\alpha,j\beta} = UNn_0\varphi_{i\alpha}^2
\delta_{ij}\delta_{\alpha\beta}$, $[\mu_{\rm eff}]_{i\alpha,j\beta} = (\mu-2UNn_0|\phi_{i\alpha}|^2)\delta_{i\alpha,j\beta}$, and $N_{\rm b}$, $N_{\rm c}$ are the
number of bands and unit cells, respectively. The condensate fraction
is defined as $n_0=N_0/N$,
where $N_0$ is the number of condensed particles and $N=N_{\rm
  b}N_{\rm c}$ is the total number of sites.

\subsection{Minimization of $E_{\rm MF}$}

The state condensation occurs in, $\ket{\varphi_0}$, minimizes the
mean-field energy $E_{\rm MF}$, with $\mu$ a Lagrange multiplier enforcing the
normalization $\braket{\varphi_0}{\varphi_0}=1$. In other words, we
need to find $\varphi_{i\alpha}$, $\varphi_{i\alpha}^*$ so that
\begin{equation}
  \begin{aligned}
  E_{\rm MF} =& n_0\sum_{i\alpha,j\beta} \varphi_{i\alpha}^* [H_{\rm
      kin}]_{i\alpha,j\beta} \varphi_{j\beta} - n_0\mu
  \sum_{i\alpha}\varphi_{i\alpha}^* \varphi_{i\alpha} \\
  +& N\frac{Un_0^2}{2}\sum_{i\alpha}(\varphi_{i\alpha}^*)^2\varphi_{i\alpha}^2,
  \end{aligned}
\end{equation}
is minimized, meaning that $\varphi_{i\alpha}$ satisfies
\begin{equation}
  n_0\sum_{j\beta}[H_{\rm kin}]_{i\alpha,j\beta}\varphi_{j\beta} +
  NUn_0^2|\varphi_{i\alpha}|^2\varphi_{i\alpha}-\mu n_0\varphi_{i\alpha} = 0. \label{eq.wf_l_equation}
\end{equation}
From these relations, we obtain for
$\braket{\varphi_0}{\varphi_0}=1$, 
\begin{equation}
  \mu = \bra{\varphi_0}H_{\rm kin}\ket{\varphi_0} +
  NUn_0\sum_{i\alpha}|\varphi_{i\alpha}|^4. 
\end{equation}

When $\ket{\varphi_0}$ minimizes
$E_{\rm MF}$, we can verify that 
$\gamma_zH_B(\ket{\varphi_0},-\ket{\varphi_0^*})^T=0$. Defining the matrix
$[F]_{i\alpha,j\beta} = |\varphi_{i\alpha}|^2\delta_{i\alpha,j\beta}$,
\begin{equation}
  \begin{aligned}
    &\gamma_zH_B\begin{pmatrix}
    \ket{\varphi_0}\\
    -\ket{\varphi_0^*}
    \end{pmatrix}\\ =& \begin{pmatrix}
      H_{\rm kin}\ket{\varphi_0} - \mu\ket{\varphi_0} +
      2Un_0F\ket{\varphi_0} - \Delta\ket{\varphi_0^*} \\
      H_{\rm kin}^*\ket{\varphi_0^*} - \mu\ket{\varphi_0^*} +
      2Un_0F\ket{\varphi_0^*} - \Delta^{\dag}\ket{\varphi_0}
    \end{pmatrix}. \label{eq.vanishing}
  \end{aligned}
\end{equation}
Since $\Delta\ket{\varphi_0^*} = Un_0F\ket{\varphi_0}$, the first
component can be rewritten as $(H_{\rm kin} +
Un_0F)\ket{\varphi_0}-\mu\ket{\varphi_0}$.
On the other hand, Eq.~\eqref{eq.wf_l_equation} can be rewritten
as $(H_{\rm kin}+Un_0F)\ket{\varphi_0} = \mu\ket{\varphi_0}$. The first
component therefore vanishes.  Similarly, the second component of
Eq.~\eqref{eq.vanishing} also vanishes.

In the main text, we assume $|\varphi_{i\alpha}|=1/\sqrt{N}$ is
uniform, and that $\ket{\varphi_0}$ is a flat-band eigenstate with
energy $\varepsilon_0$. Then
\begin{equation}
  \begin{aligned}
    \mu &= \varepsilon_0 + Un_0,\\
    \mu_{\rm eff} &= \varepsilon_0 - Un_0.
  \end{aligned}
\end{equation}
Note that these expressions coincide with those used in
Refs.~\onlinecite{Julku2021a,Julku2021b,Julku2023}, since we define $n_0$
as the number of condensed particles per site. 

\subsection{Paraunitary standardization and zero-point energy}

To determine the eigenmodes of our Hamiltonian, we seek to write
$\crea{\vec{\psi}}H_B\anni{\vec{\psi}} =
\crea{\vec{\gamma}}\vec{\varepsilon}\anni{\vec{\gamma}}$, where
$\crea{\vec{\gamma}} =
(\crea{\gamma_{1+}},\ldots,\crea{\gamma_{N+}},\anni{\gamma_{1-}},\ldots,\anni{\gamma_{N-}})$
contains Bogoliubov boson operators, and $\vec{\varepsilon}$ is
diagonal or at least of a form which makes the spectrum clear. To this
end, we need to find $\Gamma$ and $\vec{\varepsilon}$ such that
$H_B=\Gamma^{\dag}\vec{\varepsilon}\Gamma$, meaning that $\vec{\gamma}
= \Gamma\vec{\psi}$. Because the operators $\vec{\gamma}$ need to
fulfill bosonic commutation relations, we need
$\Gamma\gamma_z\crea{\Gamma} = \gamma_z$, where $\gamma_z ={\rm diag}
\sigma_z\otimes \mathbb{1}_N$:
\begin{equation}
  \begin{aligned}
    [[\anni{\vec{\gamma}}]_i,[\crea{\vec{\gamma}}]_j] =&
    [\gamma_z]_{ij} \\
    =& \sum_{kl}
    [\Gamma_{ik}[\anni{\vec{\psi}}]_k,[\crea{\psi}]_l\Gamma_{jl}^*] \\
    =& \sum_{kl} \Gamma_{ik}[\gamma_z]_{kl}[\crea{\Gamma}]_{lj}. 
  \end{aligned}
\end{equation}
In other words, in contrast to fermionic Hamiltonians where it is
sufficient to diagonalize $H_B$, we need to carry out paraunitary
diagonalization (or at least
standardization)~\cite{Colpa1978,Colpa1986a,Colpa1986b}. Now notice
that, when $\Gamma\gamma_z\crea{\Gamma} = \gamma_z$, 
\begin{equation}
    H_B = \crea{\Gamma}\vec{\varepsilon}\anni{\Gamma} 
    \Leftrightarrow \Gamma \gamma_zH_B\Gamma^{-1} = \gamma_z\vec{\varepsilon},
\end{equation}
i.e. we can obtain $\Gamma$ and $\vec{\varepsilon}$ by diagonalizing
$\gamma_zH_B$, provided it is diagonalizable by a matrix fulfilling
$\Gamma\gamma_z\crea{\Gamma} = \gamma_z$.

In Bogoliubov theory, this is unfortunately not quite the case: the
matrix $\gamma_zH_B$ has a vanishing eigenvalue with a geometric
multiplicity of one and an algebraic multiplicity of two, meaning it
is not diagonalizable. In this case, it is impossible to find a
diagonal $\vec{\varepsilon}$ while keeping the components of $\Gamma$
finite. However, it is possible to keep $\Gamma$ finite if we allow
$\vec{\varepsilon}$ to have the nondiagonal standard
form~\cite{Colpa1986a,Colpa1986b}:
\begin{equation}
  \vec{\varepsilon} = \begin{pmatrix}
    \vec{\varepsilon_{>0}} & 0 & 0 & 0 \\
    0 & \mathbb{1}_{n_{\rm zeroes}} & 0 & J \\
    0 & 0 & \vec{\varepsilon_{>0}} & 0 \\
    0 & J & 0 & \mathbb{1}_{n_{\rm zeroes}}
  \end{pmatrix},
\end{equation}
where $\vec{\varepsilon_{>0}}$ is a diagonal matrix containing the
eigenvalues with equal geometric and algebraic multiplicity (in our
case, nonzero eigenvalues) and $n_{\rm zeros}$ is the number of
improper zero modes (in Bogoliubov theory, $n_{\rm zeros}=1$). The
matrix $J$ is diagonal with $[J]_{ii}\equiv \eta_{i}$, with each
$\eta_i$ either $1$ or $-1$.

Hamiltonians in this standardized form have a known
spectrum~\cite{Colpa1986b}, and the zero-point energy of
$\crea{\vec{\psi}}H_B\anni{\vec{\psi}}$ coincides with the zero-point
energy of
\begin{equation}
  \begin{aligned}
    &\sum_{i=1}^{N-n_{\rm
        zeros}} \varepsilon_{i} (\crea{\gamma_{i+}}\anni{\gamma_{i+}} +
    \anni{\gamma_{i-}}\crea{\gamma_{i-}} ) \\
    =& \sum_{i=1}^{N-n_{\rm
        zeros}} \varepsilon_{i} (\crea{\gamma_{i+}}\anni{\gamma_{i+}}+
    \crea{\gamma_{i-}}\anni{\gamma_{i-}}) 
    + \sum_{i=1}^{N-n_{\rm
        zeros}} \varepsilon_i.
  \end{aligned}
\end{equation}

The zero-point energy for the Hamiltonian
$\crea{\vec{\psi}}H_B\anni{\vec{\psi}}$ is therefore
$\sum_{i=1}^{N-n_{\rm zeroes}}\varepsilon_i$.

The full Hamiltonian considered in this work is
\begin{equation}
  \frac{H}{N} = \frac{1}{2N}\crea{\vec{\psi}}H_B\anni{\vec{\psi}} +
  E_{\rm MF} + E_0. 
\end{equation}
Since $\ket{\varphi_0}$ minimizes the mean-field energy
$E_{\rm MF}$ and is a flat-band eigenstate, $E_{\rm MF}$ and $E_0$ are
independent of the particular $\ket{\varphi_0}$. We therefore define
the zero-point energy as
\begin{equation}
  \Lambda = \frac{1}{2N}\sum_n\varepsilon_n,
\end{equation}
where the sum runs over the strictly positive eigenvalues of
$\gamma_zH_B$. This is the definition used in the main text.

\subsubsection{Numerical diagonalization of $\gamma_zH_B$}

The diagonalization of $\gamma_zH_B$ can in principle be performed
directly, with the caveat that it is not diagonalizable on the
eigenspace of zero modes. However, especially in the presence of
degenerate nonzero eigenvalues, the eigenvectors might require further
processing to fulfill the required condition
$\Gamma\gamma_z\crea{\Gamma} = \gamma_z$. To circumvent this issue, we
employ the method proposed in Ref.~\onlinecite{Colpa1978}, which
requires no diagonalization of non-Hermitian matrices and always
provides a $\Gamma$ fulfilling $\Gamma\gamma_z\crea{\Gamma} =
\gamma_z$. Reference~\onlinecite{Colpa1978} mostly covers cases where
$H_B$ is positive definite, which is not the case here. However,
treating a positive semidefinite $H_B$ as the limit of a positive
definite matrix allows us to obtain eigenvectors corresponding to
eigenvalues which remain nonzero in this limit. We do not need
knowledge of the zero modes, but they can be determined using an
algorithm proposed in Refs.~\cite{Colpa1986a,Colpa1986b}.

When eigenvectors are needed, we use the following procedure:
\begin{enumerate}
  \item { We add a small positive shift to make $H_B$ positive
    definite, $\widetilde{H_B}= H_B+\epsilon \mathbb{1}_{2N}$. This
    allows us to determine a Cholesky decomposition of
    $\widetilde{H_B}=\mathcal{K}^{\dag}\mathcal{K}$.}
  \item{ We construct the Hermitian matrix $\widetilde{G} =
    \mathcal{K}\gamma_z\mathcal{K}^{\dag}$, and diagonalize it to obtain
    $\widetilde{G} = \mathcal{U}\mathcal{L}\mathcal{U}^{\dag}$, where
    $\mathcal{U}$ is a unitary matrix and $\mathcal{L}$ is a diagonal
    matrix with an equal number of positive and negative entries. We
    order the entries in $\mathcal{L}$ so that the first $N$ are
    positive, and the last $N$ are negative. We then take 
    $\vec{\varepsilon}=\gamma_z\mathcal{L}$, which only contains
    positive entries.}
    \item{ We now take $\Gamma =
      (\gamma_z\mathcal{L})^{-1/2}\mathcal{U}^{\dag} \mathcal{K}$. We
      can verify that
      \begin{equation}
        \begin{aligned}
          \Gamma\gamma_z\Gamma^{\dag} =&
          (\gamma_z\mathcal{L})^{-1/2}
          \mathcal{U}^{\dag}
          \mathcal{K}
          \gamma_z
          \mathcal{K}^{\dag}
          \mathcal{U}
          (\gamma_z\mathcal{L})^{-1/2} \\
          =& (\gamma_z\mathcal{L})^{-1/2}
          \mathcal{U}^{\dag}
          (\mathcal{U}\mathcal{L}\mathcal{U}^{\dag})
          \mathcal{U}
          (\gamma_z\mathcal{L})^{-1/2} \\
          =& \gamma_z, \\
          \Gamma^{\dag}\vec{\varepsilon}\Gamma =&
          \mathcal{K}^{\dag}
          \mathcal{U}(\gamma_z\mathcal{L})^{-1/2}(\gamma_z\mathcal{L})
          (\gamma_z\mathcal{L})^{-1/2} \mathcal{U}^{\dag}\mathcal{K}
          \\
          =& \mathcal{K}^{\dag}\mathcal{K} = H_B + \epsilon\mathbb{1}_{2N}. 
        \end{aligned}
      \end{equation}
    }
\end{enumerate}

By taking $\epsilon$ small enough, we can obtain the eigenvalues and
eigenvectors corresponding to nonzero eigenvalues of $\gamma_z H_B$
using this method, without diagonalizing a non-hermitian
matrix. However, this method often turns out to be slower than the
direct diagonalization of $\gamma_zH_B$, and we therefore use it only
when eigenvectors are required.

\subsection{Computation of $n_0$}

We compute $n_0$ self-consistently. We start from an initial guess for
$n_0$, and construct the Hamiltonian $H_B$ with this value. We then
evaluate the excitation fraction $n_{\rm ex}$, and use that $n_{\rm
  tot} = n_0+n_{\rm ex}$ to obtain a new value for $n_0$. We 
reconstruct $H_B$ with this new value and repeat the procedure until
$n_0$ has converged. Note that $\ket{\varphi_0}$ is not updated during
this procedure, since on a flat band the set of wavefunctions
minimizing the mean-field energy does not depend on $n_0$.

The excitation fraction can be computed from the matrices
$\vec{\varepsilon}$ and $\Gamma$ obtained as a result of the
paraunitary standardization:
\begin{equation}
  \begin{aligned}
    n_{\rm ex} =& \frac{1}{N} \sum_{i\alpha}'
    \ave{\crea{c_{i\alpha}}\anni{c_{i\alpha}}} = \frac{1}{2N}\sum_{i\alpha}' \left(
    \ave{\crea{c_{i\alpha}}\anni{c_{i\alpha}}} +
    \ave{\anni{c_{i\alpha}}\crea{c_{i\alpha}}} - 1 \right) \\
    =& \frac{1}{2N}\sum_{i}'\left(
    \ave{[\crea{\vec{\psi}}]_i,[\anni{\vec{\psi}}]_i} - 1 \right) \\
    =& \frac{1}{2N} \sum_{ijk}' \left(
    [(\Gamma^{-1})^{\dag}]_{ji}[\Gamma^{-1}]_{ik}
    \ave{[\crea{\vec{\gamma}}]_j[\anni{\vec{\gamma}}]_k} - 1 \right),
  \end{aligned}
\end{equation}
where $\sum'$ indicates that the zero modes should be excluded from
the summation. 

For a more convenient notation, we can define $\ket{\psi_{n+}}$ the
$n$:th column of $\Gamma^{-1}$ and $\ket{\psi_{n-}}$ the $n+N$:th
column. Then, since the Bogoliubov bosons follow bosonic statistics,
we obtain at zero temperature that
\begin{equation}
  n_{\rm ex} = \frac{1}{2N}\sum_{n=1}^{N-n_{\rm zeroes}}\left( -1+
  \braket{\psi_{n-}}{\psi_{n-}} \right).   
\end{equation}
To determine $n_0$, we therefore need the eigenvectors corresponding
to strictly negative eigenvalues of $\gamma_zH_B$.

\section{Role of non-contractible loop states}

\begin{figure}
  \includegraphics[width=\columnwidth]{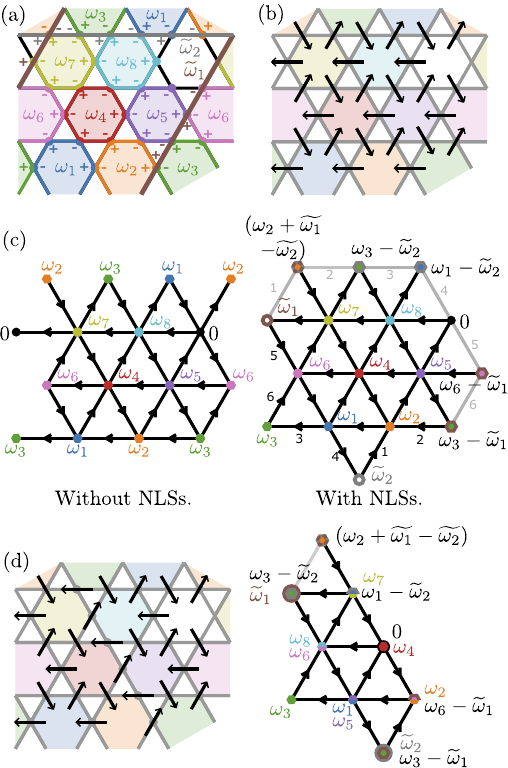}
  \caption{(a) Possible choice of basis of CLSs and NLSs in a kagome
    lattice with periodic boundary conditions and $3\times 3$ unit
    cells. (b) Wavefunction corresponding to the $\vec{k}=(0,0)$ Bloch
    function. Each arrow represents the complex number
    $\varphi_{i\alpha}$ and has length $1/\sqrt{N}$. (c) Left:
    Coefficients $\omega_i$ corresponding to the wavefunction in (b)
    when NLSs are not included. Some points, such as $\omega_1$,
    $\omega_2$, $\omega_3$ and even $0$ would need to be at several
    distinct positions in the complex plane to reproduce the desired
    phase pattern: the wavefunction in (b) can not be constructed from
    CLSs alone. Right: Configuration of coefficients when NLSs are
    also included. In this case, coefficients are well-defined, and
    some vertices of the framework correspond to linear combination of
    the various coefficients instead of the coefficients
    themselves. Note that because of periodic boundary conditions, the
    numbered black edges need to match the direction of the
    corresponding numbered gray edges. (d) Example of another
    candidate wavefunction, which is not a Bloch function, and can
    only be constructed by allowing for nonzero $\widetilde{\omega}_1$
    and $\widetilde{\omega}_2$.} \label{fig.nlss}
\end{figure}

In the main text, we considered constraints imposed by the overlap of
CLSs at sites which do not involve the NLSs in our basis. As the
system size increases, most sites will usually not be occupied by an
NLS, but they play an important role in systems with periodic boundary
conditions. To illustrate this, we consider a $3\times 3$ unit cell kagome
lattice with periodic boundary conditions. 

In the kagome lattice, the overlaps, or absence thereof, at the sites
which do not involve an NLS (see Fig.~\ref{fig.nlss}a) impose the
following constraints for a uniform $|\varphi_{i\alpha}|$:
\begin{enumerate}
  \item $|\omega_5|=|\omega_8|=1/\sqrt{N}$ from the sites which do not
    have overlaps due to the plaquette which has no associated CLS
    (the CLS on this plaquette can be constructed from the other CLSs
    and NLSs, and is therefore not needed),
  \item $|\omega_i-\omega_j| = 1/\sqrt{N}$ for the pairs $\{i,j\} =
    \{1,2\}$, $\{1,3\}$, $\{1,4\}$, $\{1,6\}$, $\{2,4\}$, $\{2,5\}$, $\{3,6\}$,
    $\{4,5\}$, $\{4,6\}$, $\{4,7\}$, $\{4,8\}$, $\{5,8\}$, $\{6,7\}$, $\{7,8\}$.  
\end{enumerate}

From the sites which involve NLSs, on the other hand, we get
\begin{enumerate}
  \setcounter{enumi}{2}
\item
  $|\omega_1-\widetilde{\omega_2}|=|\omega_2-\widetilde{\omega_2}|=|\omega_6-\widetilde{\omega_1}| 
  = |\omega_7-\widetilde{\omega_1}|
  = 1/\sqrt{N}$ from sites surrounding the plaquette without an
  associated CLS, 
\item
  $|\omega_2+\widetilde{\omega_1}-\omega_7-\widetilde{\omega_2}|=1/\sqrt{N}$
  from the site where the two NLSs cross,
\item $|\omega_i-\omega_j-\widetilde{\omega_1}| = 1/\sqrt{N}$ for
  $\{i,j\} = \{2,3\}$, $\{5,3\}$ ,$\{5,6\}$, and
  $|\omega_i-\omega_j-\widetilde{\omega_2}| = 1/\sqrt{N}$ for
  $\{i,j\}=\{3,7\}$, $\{3,8\}$, $\{1,8\}$ from the other sites. 
\end{enumerate}

In total, this gives 27 constraints on the positions of $\omega_i$ and
$\widetilde{\omega}_i$ in the complex plane, one for each site. In the
absence of NLSs, i.e. setting $\widetilde{\omega}_i=0$, the Bloch
function corresponding to $\vec{k}=(0,0)$, shown in
Fig.~\ref{fig.nlss}(b), could not be constructed. However, this is
possible to achieve once NLSs are also included (see
Fig.~\ref{fig.nlss}(c)). This is expected, since
$\vec{k}=(0,0)$ has a singular band touching, which necessitates the
inclusion of NLSs.  

If NLSs are not included, problems can occur also for states that are
not simply Bloch functions but involve the $\vec{k}=(0,0)$, as shown
in Fig.~\ref{fig.nlss}(d).

\section{Verifying stability without explicitely considering
  frameworks} \label{sec.no-constraints}

A particular $\ket{\varphi_0}$ is a flat-band eigenstate if and only if
\begin{equation}
  \ket{\varphi_0} = T\vec{\omega}, \label{eq.cls_to_sites}
\end{equation}
where $\vec{\omega}=(\omega_1,\ldots,\omega_{n_{\rm
    CLS}},\widetilde{\omega}_1,\ldots,\widetilde{\omega}_{n_{\rm
    NLS}})$, with $n_{\rm CLS}$ and $n_{\rm NLS}$ the numbers of
CLSs and NLSs, respectively. The matrix $T$ is of dimension
$N\times(n_{\rm CLS}+n_{\rm NLS})$, with $[T]_{i\alpha,j} =
\nu_{(\vec{R}_j-\vec{R}_i)\alpha}$, $j=1,\ldots,n_{\rm CLS}$, the amplitude of the
$j$:th CLS at site $i\alpha$, and
$[T]_{i\alpha,j+n_{\rm CLS}}=\widetilde{\nu}_{i\alpha,j}$,
$j=1,\ldots,n_{\rm NLS}$ the amplitude of the
$j$:th NLS at site $i\alpha$. 

For a particular candidate wavefunction $\ket{\varphi_0}$,
Eq.~\eqref{eq.cls_to_sites} has nontrivial solutions
$\vec{\omega}$ if and only if
$TT^+\ket{\varphi_0}=\ket{\varphi_0}$, where $T^+$ is the
Moore-Penrose pseudoinverse of $T$. A nontrivial solution
$\vec{\omega}$ also exists if and only if $\ket{\varphi_0}$ can be
constructed from the CLSs and NLSs, meaning it is a flat-band
eigenstate. We assume this is the case for
some $\ket{\varphi_0}$ with uniform $|\varphi_{i\alpha}|$.

A stable condensate is impossible in $\ket{\varphi_0}$ if there exists
a pair of flat-band states $C\ket{\varphi_0}$,
$C^{\dag}\ket{\varphi_0}$ where $C$ is a diagonal matrix (or a single
flat-band state $C\ket{\varphi_0}$ when $C$ is real). In order for
stable condensation to be possible, there should therefore exist no
$C$ not proportional to identity such that
$TT^+C\ket{\varphi_0}=C\ket{\varphi_0}$ and
$TT^+C^{\dag}\ket{\varphi_0}=C^{\dag}\ket{\varphi_0}$
simultaneously.

Let us now define a diagonal unitary matrix $[A]_{i\alpha,j\beta} =
e^{i\theta_{i\alpha}}\delta_{i\alpha,j\beta}$, so that
$\ket{\varphi_0} = A\ket{u}$, with $\ket{u} = 
(1,\ldots,1)/\sqrt{N}$ a real vector. Then $G\ket{u}=\ket{u}$, where
$G=A^{\dag}TT^+A$ is a Hermitian matrix. If a pair of states
$C\ket{\varphi_0}$ and $C^{\dag}\ket{\varphi_0}$ exists on the flat band,
$G\ket{v}=\ket{v}$ and $G\ket{v^*}=\ket{v^*}$, where
$\ket{v}=C\ket{u}$, i.e. $G$ has a corresponding
eigenvector corresponding to a unit eigenvalue such that its complex
conjugate is also an eigenvector of $G$ corresponding to a unit
eigenvalue. Conversely, if there exists a vector $\ket{x}$ such that
$G\ket{x}=\ket{x}$ and $G\ket{x^*}=\ket{x^*}$ simultaneously, we can
define $[U]_{i\alpha,i\alpha}={\rm arg}[\ket{x}]_{i\alpha}$ and
$[R]=\sqrt{N}|[\ket{x}]_{i\alpha}|$ to recover a pair of states
$A\ket{x}=RU\ket{\varphi_0}$ and $A\ket{x^*} =
RU^{\dag}\ket{\varphi_0}$. A pair of states $C\ket{\varphi_0}$,
$C^{\dag}\ket{\varphi_0}$ therefore exists on the flat band if and
only if $G$ has a corresponding eigenvector with unit eigenvalue such that its
complex conjugate is also an eigenvector of $G$ with unit eigenvalue. 

Such vectors are simultaneous eigenvectors of $G$ and $G^*$. For
stable condensation to be possible, we therefore need $G$ and $G^*$ to
have exactly one simultaneous eigenvector corresponding to a unit
eigenvalue, $\ket{u}$. In other words, the intersection $V =
{\rm ker}[G-\mathbb{1}]\bigcap {\rm   ker}[G^*-\mathbb{1}]$ should be
one-dimensional for stable condensation to be possible. 

Note that for a large system, because of the translational invariance,
it is usually sufficient to consider a system of fewer unit cells
to verify that the set of CLSs and NLSs is compatible with a stable BEC. 

\subsection{Example: kagome lattice}

\begin{figure}
  \includegraphics[width=\columnwidth]{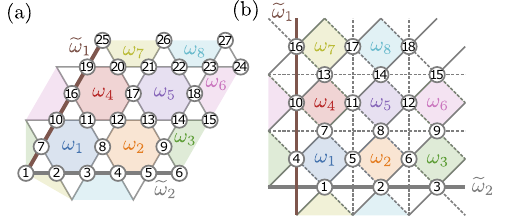}
  \caption{Numbering of sites and CLSs/NLSs used in the (a) kagome and
  (b) checkerboard lattice when building the matrix $T$ (see
    Eq.~\eqref{eq.cls_to_sites}). } \label{fig.numbering}
\end{figure}

In the kagome lattice, using the numbering shown in
Fig.~\ref{fig.numbering}(a), a possible matrix $T$ for a system of
$3\times 3$ unit cells is
\begin{equation}
  T = \begin{pmatrix}
    \g{0} & \g{0} & -1 & \g{0} & \g{0} & \g{0} & +1 & \g{0} & -1 & +1 \\
    +1 & \g{0} & \g{0} & \g{0} & \g{0} & \g{0} & -1 & \g{0} & \g{0} & -1 \\
    -1 & \g{0} & \g{0} & \g{0} & \g{0} & \g{0} & \g{0} & +1 & \g{0} & +1 \\
    \g{0} & +1 & \g{0} & \g{0} & \g{0} & \g{0} & \g{0} & -1 & \g{0} & -1 \\
    \g{0} & -1 & \g{0} & \g{0} & \g{0} & \g{0} & \g{0} & \g{0} & \g{0} & +1 \\
    \g{0} & \g{0} & +1 & \g{0} & \g{0} & \g{0} & \g{0} & \g{0} & \g{0} & -1 \\
    -1 & \g{0} & +1 & \g{0} & \g{0} & \g{0} & \g{0} & \g{0} & +1 & \g{0} \\
    +1 & -1 & \g{0} & \g{0} & \g{0} & \g{0} & \g{0} & \g{0} & \g{0} & \g{0} \\
    \g{0} & +1 & -1 & \g{0} & \g{0} & \g{0} & \g{0} & \g{0} & \g{0} & \g{0} \\
    +1 & \g{0} & \g{0} & \g{0} & \g{0} & -1 & \g{0} & \g{0} & -1 & \g{0} \\
    -1 & \g{0} & \g{0} & +1 & \g{0} & \g{0} & \g{0} & \g{0} & \g{0} & \g{0} \\
    \g{0} & +1 & \g{0} & -1 & \g{0} & \g{0} & \g{0} & \g{0} & \g{0} & \g{0} \\
    \g{0} & -1 & \g{0} & \g{0} & +1 & \g{0} & \g{0} & \g{0} & \g{0} & \g{0} \\
    \g{0} & \g{0} & +1 & \g{0} & -1 & \g{0} & \g{0} & \g{0} & \g{0} & \g{0} \\
    \g{0} & \g{0} & -1 & \g{0} & \g{0} & +1 & \g{0} & \g{0} & \g{0} & \g{0} \\
    \g{0} & \g{0} & \g{0} & -1 & \g{0} & +1 & \g{0} & \g{0} & +1 & \g{0} \\
    \g{0} & \g{0} & \g{0} & +1 & -1 & \g{0} & \g{0} & \g{0} & \g{0} & \g{0} \\
    \g{0} & \g{0} & \g{0} & \g{0} & +1 & -1 & \g{0} & \g{0} & \g{0} & \g{0} \\
    \g{0} & \g{0} & \g{0} & +1 & \g{0} & \g{0} & \g{0} & \g{0} & -1 & \g{0} \\
    \g{0} & \g{0} & \g{0} & -1 & \g{0} & \g{0} & +1 & \g{0} & \g{0} & \g{0} \\
    \g{0} & \g{0} & \g{0} & \g{0} & +1 & \g{0} & -1 & \g{0} & \g{0} & \g{0} \\
    \g{0} & \g{0} & \g{0} & \g{0} & -1 & \g{0} & \g{0} & +1 & \g{0} & \g{0} \\
    \g{0} & \g{0} & \g{0} & \g{0} & \g{0} & +1 & \g{0} & -1 & \g{0} & \g{0} \\
    \g{0} & \g{0} & \g{0} & \g{0} & \g{0} & -1 & \g{0} & \g{0} & \g{0} & \g{0} \\
    \g{0} & \g{0} & \g{0} & \g{0} & \g{0} & \g{0} & -1 & \g{0} & +1 & \g{0} \\
    \g{0} & \g{0} & \g{0} & \g{0} & \g{0} & \g{0} & +1 & -1 & \g{0} & \g{0} \\
    \g{0} & \g{0} & \g{0} & \g{0} & \g{0} & \g{0} & \g{0} & +1 & \g{0} & \g{0} \\
  \end{pmatrix}
\end{equation}

We can verify for instance the stability of condensation into the
$\vec{k}=(0,0)$ state
\begin{equation}
  \begin{aligned}
  \ket{\varphi_0} = \frac{1}{\sqrt{N}}\big(&
    e^{-i\frac{\pi}{3}} ,
    e^{i\frac{\pi}{3}} ,
    e^{-i\frac{\pi}{3}} ,
    e^{i\frac{\pi}{3}} ,
    e^{-i\frac{\pi}{3}} ,
    e^{i\frac{\pi}{3}} ,
    -1 ,
    -1 ,
    -1 ,\\
    &e^{-i\frac{\pi}{3}} ,
    e^{i\frac{\pi}{3}} ,
    e^{-i\frac{\pi}{3}} ,
    e^{i\frac{\pi}{3}} ,
    e^{-i\frac{\pi}{3}} ,
    e^{i\frac{\pi}{3}} ,
    -1 ,
    -1 ,
    -1 ,\\
    &e^{-i\frac{\pi}{3}} ,
    e^{i\frac{\pi}{3}} ,
    e^{-i\frac{\pi}{3}} ,
    e^{i\frac{\pi}{3}} ,
    e^{-i\frac{\pi}{3}} ,
    e^{i\frac{\pi}{3}} ,
    -1 ,
    -1 ,
    -1 )^T.
    \end{aligned}
\end{equation}
We numerically solve for $V={\rm
  ker}[G-\mathbb{1}]\bigcap {\rm ket}[G^*-\mathbb{1}]$, and find that
it is one-dimensional (spanned by a vector with all entries equal),
meaning that for this $\ket{\varphi_0}$, the condensate can be stable. 

\subsection{Example: checkerboard lattice}

In the checkerboard lattice considered in the main text, we again
consider a $3\times 3$ unit cell system, and use the numbering shown in
Fig.~\ref{fig.numbering}(b) to obtain
\begin{equation}
  T = \begin{pmatrix}
    -1 & \g{0} & \g{0} & \g{0} & \g{0} & \g{0} & +1 & \g{0} & +1 & \g{0} \\
    \g{0} & -1 & \g{0} & \g{0} & \g{0} & \g{0} & \g{0} & +1 & +1 & \g{0} \\
    \g{0} & \g{0} & -1 & \g{0} & \g{0} & \g{0} & \g{0} & \g{0} & +1 & \g{0} \\
    -1 & \g{0} & +1 & \g{0} & \g{0} & \g{0} & \g{0} & \g{0} & \g{0} & +1 \\
    +1 & -1 & \g{0} & \g{0} & \g{0} & \g{0} & \g{0} & \g{0} & \g{0} & \g{0} \\
    \g{0} & +1 & -1 & \g{0} & \g{0} & \g{0} & \g{0} & \g{0} & \g{0} & \g{0} \\
    +1 & \g{0} & \g{0} & -1 & \g{0} & \g{0} & \g{0} & \g{0} & \g{0} & \g{0} \\
    \g{0} & +1 & \g{0} & \g{0} & -1 & \g{0} & \g{0} & \g{0} & \g{0} & \g{0} \\
    \g{0} & \g{0} & +1 & \g{0} & \g{0} & -1 & \g{0} & \g{0} & \g{0} & \g{0} \\
    \g{0} & \g{0} & \g{0} & -1 & \g{0} & +1 & \g{0} & \g{0} & \g{0} & +1 \\
    \g{0} & \g{0} & \g{0} & +1 & -1 & \g{0} & \g{0} & \g{0} & \g{0} & \g{0} \\
    \g{0} & \g{0} & \g{0} & \g{0} & +1 & -1 & \g{0} & \g{0} & \g{0} & \g{0} \\
    \g{0} & \g{0} & \g{0} & +1 & \g{0} & \g{0} & -1 & \g{0} & \g{0} & \g{0} \\
    \g{0} & \g{0} & \g{0} & \g{0} & +1 & \g{0} & \g{0} & -1 & \g{0} & \g{0} \\
    \g{0} & \g{0} & \g{0} & \g{0} & \g{0} & +1 & \g{0} & \g{0} & \g{0} & \g{0} \\
    \g{0} & \g{0} & \g{0} & \g{0} & \g{0} & \g{0} & -1 & \g{0} & \g{0} & +1 \\
    \g{0} & \g{0} & \g{0} & \g{0} & \g{0} & \g{0} & +1 & -1 & \g{0} & \g{0} \\
    \g{0} & \g{0} & \g{0} & \g{0} & \g{0} & \g{0} & \g{0} & +1 & \g{0} & \g{0} 
  \end{pmatrix}
\end{equation}

A flat-band eigenstate with uniform $|\varphi_{i\alpha}|$ is
\begin{equation}
  \begin{aligned}
    \ket{\varphi_0} = \big(&1, 1, 1, i, i, i, \\
    &1, 1, 1, i, i, i, \\
    &1, 1, 1, i, i, i\big)^T.
  \end{aligned}
\end{equation}

In this case, $V$ is six-dimensional, spanned by six real vectors
\begin{equation}
  \begin{aligned}
    v_1 =& (1,1,1,\g{0},\g{0},\g{0},\g{0},\g{0},\g{0},\g{0},\g{0},\g{0},\g{0},\g{0},\g{0},\g{0},\g{0},\g{0})^T, \\
    v_2 =& (\g{0},\g{0},\g{0},\g{0},\g{0},\g{0},1,1,1,\g{0},\g{0},\g{0},\g{0},\g{0},\g{0},\g{0},\g{0},\g{0})^T, \\
    v_3 =& (\g{0},\g{0},\g{0},\g{0},\g{0},\g{0},\g{0},\g{0},\g{0},\g{0},\g{0},\g{0},1,1,1,\g{0},\g{0},\g{0})^T, \\
    v_4 =& (\g{0},\g{0},\g{0},1,\g{0},\g{0},\g{0},\g{0},\g{0},1,\g{0},\g{0},\g{0},\g{0},\g{0},1,\g{0},\g{0})^T, \\
    v_5 =& (\g{0},\g{0},\g{0},\g{0},1,\g{0},\g{0},\g{0},\g{0},\g{0},1,\g{0},\g{0},\g{0},\g{0},\g{0},1,\g{0})^T, \\
    v_6 =& (\g{0},\g{0},\g{0},\g{0},\g{0},1,\g{0},\g{0},\g{0},\g{0},\g{0},1,\g{0},\g{0},\g{0},\g{0},\g{0},1)^T.
  \end{aligned}
\end{equation}
The condensate is therefore unstable. 

\subsection{Example: Tasaki lattice}

Here we consider two values of $a$ compatible with periodic boundary
conditions and a system size of $4\times 4$ unit cells: $a=\sqrt{2}$
and $a=2$. Since the matrix $T$ has dimensions $48\times 16$, we do
not provide it explicitely. In the $a=\sqrt{2}$ case, picking
$\ket{\varphi_0}$ as the Bloch function corresponding to
$\vec{k}=(\pi/2,\pi/2)$, we find that $V$ is a one-dimensional vector
space. For $a=2$, $\ket{\varphi_0}$ is the Bloch function at
$\vec{k}=(\pi,\pi)$, and $V$ turns out to be a 16-dimensional vector
space. At $a=2$, the condensate is thus unstable, while it can be
stable for $a=\sqrt{2}$.

\section{Basis-independence of the stability}

\begin{figure}
  \includegraphics[width=\columnwidth]{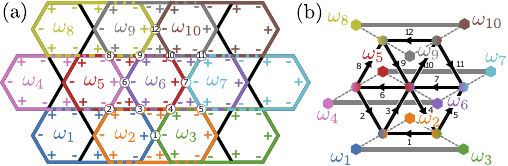}
  \caption{(a) One possible choice of CLSs in the kagome lattice. Each
    CLS occupies two horizontally adjacent hexagonal plaquettes. We do
    not show all CLSs and NLSs, but instead focus on the constraints
    imposed by the overlaps at the 12 numbered sites. (b) One
    configuration of the coefficients $\omega_i$ corresponding to
    uniform densities. The single-colored hexagons represent the
    coefficients $\omega_i$, while circles colored by gradients are
    the middle point of the dashed lines they are on (i.e. they are
    midpoints $(\omega_i+\omega_j)/2$). The thick gray
    lines have a length of $1/\sqrt{N}$. The black arrows have length
    $1/(2\sqrt{N})$, and each corresponds to $\varphi_{i}/2$, with
    $i=1\ldots 12$ one of the sites. The length of the dashed lines is
    not fixed, and the positions of $\omega_i$ can thus be
    continuously modified; however, the midpoints need to lie on a
    triangular framework, the edges of which correspond to
    $\varphi_{i}/2$.} \label{fig.other_basis}
\end{figure}

Since the approach outlined in this work relies on finding
uniform-density states $\ket{\varphi_0}$ and verifying that no states
$C\ket{\varphi_0}$, $C^{\dag}\ket{\varphi_0}$ exist on the flat band,
the results should clearly be independent of the choice of CLSs and
NLSs: no matter the basis, the states on the flat band should be the
same. Although picking a basis with more complicated CLSs can make the
constraints on $\omega_i$ more difficult to interpret geometrically,
the choice of CLSs indeed does not matter for conclusions regarding
possible $\ket{\varphi_0}$ and whether condensation is possible in
them.

As an example, let us consider the kagome lattice. In the main text,
we showed that all uniform-density states can be represented by a
triangular framework, with each $\varphi_{i\alpha}$ corresponding to
one edge of an equilateral triangle. Let us now consider a different
choice of CLSs, and show that it leads to the same conclusion. Instead
of CLSs occupying one hexagonal plaquette each, we choose CLSs
occupying two adjacent plaquettes (see
Fig.~\ref{fig.other_basis}(a)). Because of this choice, there are now
sites at which four CLSs overlap, which complicates the constraints
slightly. From the overlaps at the numbered sites in
Fig.~\ref{fig.other_basis}(a), we obtain the following constraints:
\begin{enumerate}
  \item Sites 1, 6, 7 and 12: $|\omega_i-\omega_j|=1/\sqrt{N}$ for
    $\{i,j\}=\{1,3\}$, $\{4,6\}$, $\{5,7\}$, $\{8,10\}$. This simply
    means that the coefficients in these pairs are a distance of
    $1/\sqrt{N}$ away from each other in the complex plane.
  \item {Sites 2, 3, 4, 5, 8, 9, 10 and 11:
    $|\omega_i+\omega_j-\omega_k-\omega_l|=1/\sqrt{N}$ for
    $\{i,j,k,l\} = \{1,2,4,5\}$, $\{1,2,5,6\}$, $\{2,3,5,6\}$,
    $\{2,3,6,7\}$, $\{4,5,8,9\}$, $\{5,6,8,9\}$, $\{5,6,9,10\}$,
    $\{6,7,9,10\}$. For an easier geometric interpretation, it is
    convenient to rewrite the constraint as
    \begin{equation}
      \left| \frac{\omega_i+\omega_j}{2} - \frac{\omega_k+\omega_l}{2}
      \right| = \frac{1}{2\sqrt{N}}, 
    \end{equation}
    meaning that the middle of the line between $\omega_i$ and
    $\omega_j$ is at a distance of $1/(2\sqrt{N})$ from the middle of
    the line between $\omega_k$ and $\omega_l$. 
  }
\end{enumerate}

To understand how these constraints come together, notice that
\begin{equation}
  \frac{1}{2}(\omega_i+\omega_j)-\frac{1}{2}(\omega_j+\omega_k) =
  \frac{1}{2}(\omega_i-\omega_k). 
\end{equation}
Now consider the case $i=1$, $j=2$ and $k=3$. We know the
distance between $\omega_1$ and $\omega_3$ is $1/\sqrt{N}$ by the
first set of constraints. Then the distance between
$(\omega_1+\omega_2)/2$ and $(\omega_2+\omega_3)/2$ is
$1/(2\sqrt{N})$. Moreover, since $\omega_1-\omega_3=\varphi_1$,
$(\omega_1+\omega_2)/2-(\omega_2+\omega_3)/2=\varphi_1/2$.

From the constraint from site 3, we also know that the 
distance between $(\omega_1+\omega_2)/2$ and $(\omega_5+\omega_6)/2$
is $1/(2\sqrt{N})$. Similarly, it follows from the constraint
from site 4 that the distance between $(\omega_2+\omega_3)/2$ and
$(\omega_5+\omega_6)/2$ is $1/(2\sqrt{N})$. Therefore
$(\omega_1+\omega_2)/2$, $(\omega_2+\omega_3)/2$ and
$(\omega_5+\omega_6)/2$ need to form an equilateral triangle with side
length $1/(2\sqrt{N})$. Applying a similar reasoning, we can also
deduce that the sets
\begin{equation}
  \begin{aligned}
    &\{(\omega_4+\omega_5)/2,(\omega_5+\omega_6)/2,(\omega_1+\omega_2)/2\}, \\
    &\{(\omega_5+\omega_6)/2,(\omega_6+\omega_7)/2,(\omega_2+\omega_3)/2\}, \\
    &\{(\omega_4+\omega_5)/2,(\omega_5+\omega_6)/2,(\omega_8+\omega_9)/2\}, \\
    &\{(\omega_8+\omega_9)/2,(\omega_9+\omega_{10})/2,(\omega_5+\omega_6)/2\},
    \\
    &\{(\omega_5+\omega_6)/2,(\omega_6+\omega_7)/2,(\omega_9+\omega_{10})/2\}
  \end{aligned}
\end{equation}
all form equilateral triangles in the complex plane (see
Fig.~\ref{fig.other_basis}). Each of the sides of these equilateral
triangles corresponds to $\varphi_{i}/2$ for one of the sites of the
lattice, and all possible uniform-density eigenstates are
three-colorings of the kagome lattice (there are only three distinct
orientations for the sides of the triangles, and adjacent edges of a
triangle need to have distinct orientations).

With this other choice of CLSs, the obtained set of wavefunctions is
thus the same as for the choice used in the main text, as it
should. The difference is that for the simpler choice of basis used in
the main text, each vertex of the triangular framework representing
$\ket{\varphi_0}$ corresponds to a coefficient $\omega_i$, while for
the more complicated choice used here, each vertex instead corresponds
to a linear combination $(\omega_i+\omega_j)/2$, with $i$ and $j$
corresponding to adjacent plaquettes.

\section{Stability of triangulated frameworks}

We argue here that if a flat-band state $\ket{\varphi_0}$ is
represented by a triangulated framework (i.e. each edge of a
triangulated framework corresponds to $b_{i\alpha}\varphi_{i\alpha}$),
there can exist no states $R\ket{\varphi_0}$ or pairs of states
$RU\ket{\varphi_0}$, $RU^{\dag}\ket{\varphi_0}$ on the flat band, with
$R$ a real diagonal matrix and $U$ a diagonal unitary matrix. Note
that we assume here that each triangle in the framework has nonzero
area.

Let us first consider the possibility of a pair of states
$RU\ket{\varphi_0}$, $RU^{\dag}\ket{\varphi_0}$, with $U$ a diagonal
unitary matrix. Let us start by considering a single triangle in the
triangulated framework corresponding to $\ket{\varphi_0}$. The
corresponding triangles in $RU\ket{\varphi_0}$ and
$RU^{\dag}\ket{\varphi_0}$ have identical edge lengths, because $U$
only changes angles. These two triangles are thus related by rotations
and/or reflections.

First, let us consider the case where the two triangles
are related by a single reflection. We label $1$, $2$, $3$ the corners of the
original triangle in $\ket{\varphi_0}$, $1'$, $2'$, $3'$ the corners
of the triangle in $RU\ket{\varphi_0}$ and $1''$, $2''$, $3''$ the
corners of the triangle in $RU^{\dag}\ket{\varphi_0}$. Let us focus
on the edges between $(1,2)$, $(1',2')$ and $(1'',2'')$. By
assumption, if $(1',2')$ is rotated by an angle $\theta$ compared to
$(1,2)$, $(1'',2'')$ is rotated by an angle $-\theta$ compared to
$(1,2)$. This is only possible if $(1,2)$ is parallel to the
reflection line between the triangles $(1',2',3')$ and
$(1'',2'',3'')$. The same reasoning can be used for the edges
$(1,3)$ and $(2,3)$. All three edges of the triangle $(1,2,3)$
thus need to be parallel to each other when $(1',2',3')$ and
$(1'',2'',3'')$ are related by a reflection. This goes against the
assumption of a nonzero surface area, and corresponding triangles in
$RU\ket{\varphi_0}$ and $RU^{\dag}\ket{\varphi_0}$ can not
be related by a reflection. 

Corresponding triangles in $RU\ket{\varphi_0}$,
$RU^{\dag}\ket{\varphi_0}$ are therefore related by at most an overall
rotation. Now consider an adjacent triangle to the one originally
considered. Again, the corresponding triangles in $RU\ket{\varphi_0}$
and $RU^{\dag}\ket{\varphi_0}$ need to be related by an overall
rotation. Because these triangles share an edge with the first
triangle we considered, this rotation needs to be the same as for the
first triangle. Using the same reasoning for all triangles, we get
that $RU\ket{\varphi_0}$ and $RU^{\dag}\ket{\varphi_0}$ need to be
related by an overall rotation, i.e. $U$ can only be a gauge transformation.

It remains to show that there can exist no states $R\ket{\varphi_0}$
on the flat band. Such state would correspond to a framework with the
same connectivity as $\ket{\varphi_0}$, with all edges parallel to
those in $\ket{\varphi_0}$. First, consider a single triangle in
$\ket{\varphi_0}$, and the corresponding triangle in
$\ket{\varphi_0'}=R\ket{\varphi_0}$. Since all edges of these
triangles are parallel, the two triangles are similar (assuming the
triangles have nonzero area). This means that all edges are scaled by
the same factor $c$. Now consider a triangle that shares an edge with
this first triangle: again, these triangles in $\ket{\varphi_0}$ and
$R\ket{\varphi_0}$ need to be similar. Since one edge of the triangle
in $R\ket{\varphi_0}$ is scaled by $c$ compared to the corresponding
edge in $\ket{\varphi_0}$, all edges need to be scaled by $c$. By the
same reasoning, it follows that all edges in $R\ket{\varphi_0}$ need
to be scaled by the same factor $c$ to keep the edges parallel to
those in $\ket{\varphi_0}$: $R$ is proportional to the identity and
$\ket{\varphi_0}$ and $R\ket{\varphi_0}$ are the same state up to a
gauge transformation and possibly different normalization. There can
exist no other states $R\ket{\varphi_0}$ on the flat band.

Note that the assumption of a nonzero area is essential. If a
framework consists of triangles with vanishing surface area, it
generally becomes possible to construct both pairs of states
$RU\ket{\varphi_0}$, $RU^{\dag}\ket{\varphi_0}$, and single
problematic states $R\ket{\varphi_0}$.

\section{Sawtooth ladder}

\begin{figure}
  \includegraphics[width=\columnwidth]{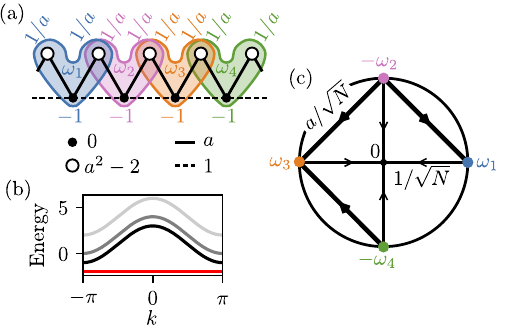}
  \caption{(a) Sawtooth ladder with a flat band at energy $-2$. A
    possible choice of CLS is shown for each unit cell (different
    colors). (b) Band dispersions for $a=1$ (black), $a=\sqrt{2}$
    (gray) and $a=2$ (light gray). The flat band of interest is shown
    in red. (c) Example of $\omega_i$ leading to uniform densities for
    $a=\sqrt{2}$. The CLSs are centered at a site
  without overlaps, which constrains $|\omega_i|=1/\sqrt{N}$ to
  achieve uniform densities. The overlaps between adjacent CLSs
  constrain $|\omega_{2n\pm
    1}+\omega_{2n}|=|\omega_{2n\pm 1}-(-\omega_{2n})|=a/\sqrt{N}$,
  constraining the distance between $\omega_{2n}$ and $\omega_{2n\pm
    1}$. As a result, $0$, $-\omega_{2n}$ and $\omega_{2n\pm 1}$ always
  form an isoceles triangle in the complex plane.}
  \label{fig.sawtooth}
\end{figure}

The sawtooth ladder, also studied in Ref.~\onlinecite{Huber2010}, is
an example of a flat-band model with CLSs compatible with bosonic
condensation. We consider a generalized version of the model with a
tuning parameter $a$ (see Fig.~\ref{fig.sawtooth}). This model always
features a lowest isolated flat band at energy $-2$, and
uniform-density eigenstates exist on the flat band for any $0<
a\leq 2$. 

As shown in Fig.~\ref{fig.sawtooth}c, the CLSs in this model overlap
in such a way that, for uniform densities, $0$, $-\omega_{2n}$ and
$\omega_{2n\pm 1}$ always need to form an isoceles triangle. Whenever
$0<a<2$, uniform-density flat-band eigenstates are then represented by
triangulated frameworks with nonzero area. This includes the usual
flat-band sawtooth model with $a=\sqrt{2}$. The area of the triangles
vanishes for $a=0$ and $a=2$. At $a=0$, the instability of
condensation is evident: the flat band is then completely trivial (it
arises from completely disconnected lattice sites). At $a=2$,
condensation is expected to be impossible due to the existence of
problematic flat-band eigenstates, similarly to the Tasaki model
considered in the main text.

\section{Tasaki flat-band model}

\subsection{Adjusted boundary conditions}

\begin{figure}
  \includegraphics[width=\columnwidth]{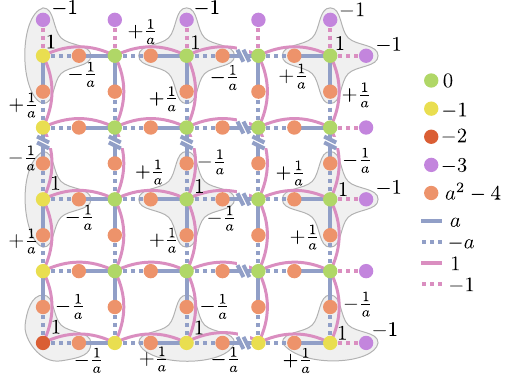}
  \caption{Tasaki model with adjusted boundary conditions. The shaded
    areas represent the different types of CLSs in the system. } \label{fig.obc}
\end{figure}

The system sizes for which we can compute zero-points energies and
$n_0$ are limited, since we need to diagonalize a $(2N)\times(2N)$
matrix. In the Tasaki lattice, uniform-density flat-band eigenstates would then be
commensurate with accessible system sizes only for a few specific
values of $a$. We therefore choose to use open boundary conditions. In
order to retain a perfectly flat band and a uniform-density $\ket{\varphi_0}$,
we modify the tight-binding parameters at the boundaries as shown in
Fig.~\ref{fig.obc}. In this case, the CLSs for unit cells in the bulk
are unchanged, but CLSs at boundaries and corners are slightly
different. However, they overlap with other CLSs in the same way as
bulk CLSs would, and the mapping explained below can be used.

\subsection{Sampling of zero-point energies}

\subsubsection{Correspondence between uniform-density flat-band eigenstates and 3-colorings}

\begin{figure}
  \includegraphics[width=\columnwidth]{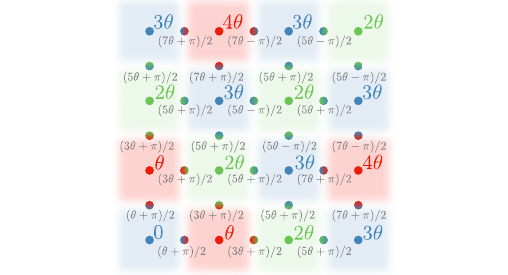}
  \caption{Mapping between three-colorings of the square lattice and
    uniform-density eigenstates $\ket{\varphi_0}$. At each site at a
    corner of a square plaquette (referred to as corner site),
    $\varphi_{i\alpha}=\omega_i$, and thus the phase increases by
    $\theta$ when moving from blue to red, red to green, or green to
    blue. Otherwise, it decreases by $\theta$. Once all $\omega_i$ are
    known, the phases at the sites on the edges of square plaquettes
    (referred to as edge sites) are easy to
    determine. } \label{fig.3_color}
\end{figure}

As mentioned in the main text, we sample uniform-density flat-band
eigenstates by taking advantage of a correspondence between them and
three-colorings of the square graph. For flat-band eigenstates with a
uniform $|\varphi_{i\alpha}|=1/\sqrt{N}$, we need
$\omega_i=e^{in\theta}/\sqrt{N}$, where $n$ is an integer and $\theta
= 2\arcsin(a/2)$, with $a\in[1,2)$ a tuning parameter. The phases of
  $\omega_i$ and $\omega_j$ corresponding to adjacent unit cells need
  to differ by exactly $\pm\theta$. We divide the possible phases of
  $\omega_i$ into three sets, $\{0,3\theta,\ldots\}$,
  $\{\theta,4\theta,\ldots\}$, $\{2\theta,5\theta,\ldots\}$, and
  associate each with a color (blue, red and green, respectively). We
  now color each unit cell with the color corresponding to the phase
  of $\omega_i$ in that unit cell. Since the phases in neighboring
  unit cells need to differ by $\pm \theta$, neighboring unit cells
  always are associated to different colors: we obtain a
  three-coloring of the square lattice.
  
We can also show that each three-coloring of a square lattice
corresponds to a uniform-density $\ket{\varphi_0}$. We start from a
three-coloring of the square lattice and associate each site of the
square lattice to a unit cell. Let us assume we know the phase
$\theta_i$ of $\omega_i$ in one unit cell $i$, and want to determine
the phase $\theta_j$ in a neighboring unit cell $j$. If the colors
associated with $i$ and $j$ are blue and red, red and green, or green
and blue, respectively ("increasing" color from $i$ to $j$)
$\theta_j=\theta_i+\theta$. Otherwise, $\theta_j=\theta_i-\theta$ (see
Fig.~\ref{fig.3_color}). As a starting point, we can choose the phase
in any unit cell, which ammounts to fixing the gauge, and then
determine the other $\omega_i$ following this procedure. This will
always yield a uniform-density wavefunction (where all neighboring cells differ
by $\pm\theta$), which can be seen by considering a group of $2\times
2$ unit cells. If we go around this plaquette, we need to start and
end at the same color. Since we work with three colors, the only
possibility is to increase the color twice and decrease the color
twice. Going around the plaquette, we thus increase the phase twice,
and decrease twice, meaning we are guaranteed to start and end with
the same phase. Since all such $2\times 2$ sets need to be consistent,
all phases will be consistent.

Note that three-colorings of the Tasaki lattice are also directly
related to three-colorings of the square lattice (we are simply adding
vertices the color of which is completely determined by the underlying
square lattice), and we could in principle interpret the three-color
sampling as three-colorings of the Tasaki lattice itself. However, it
is important to stress that the mapping to three-colorings is purely
reliant on constraints on phases of $\ket{\varphi_0}$, which are a
consequence of the way CLSs overlap, rather than of the connectivity
of the lattice. For any lattice with the same basis of CLSs as our
Tasaki model, possible $\ket{\varphi_0}$ could always be sampled from
three-colorings of the square lattice, even when the connectivity of
the actual lattice is not the same as the Tasaki lattice.

\subsubsection{Numerical results}

\begin{figure}
  \includegraphics[width=\columnwidth]{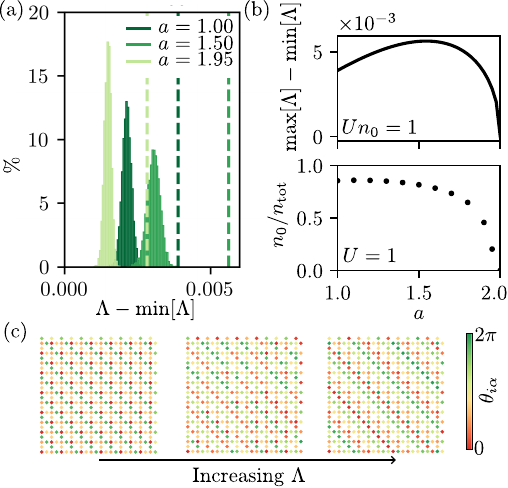}
  \caption{(a) Zero-point energies for $5000$ randomly sampled
    wavefunctions $\ket{\varphi_0}$ at different values of $a$ in the
    Tasaki model. The energies are shifted by the smallest ZPE, ${\rm
      min}[\Lambda]$. The dashed line indicates the highest ZPE, ${\rm
      max}[\Lambda]$ for each $a$. We consider a system of $12\times
    12\times 3$ sites. We set $Un_0=1$. (b) Range of ZPEs, ${\rm
      max}[\Lambda]-{\rm min}[\Lambda]$, (upper panel) and $n_0/n_{\rm
      tot}$ assuming condensation in the state with lowest ZPE (lower
    panel) and as a function of $a$. For the computation of $n_0$, we
    set $n_{\rm tot}=1$. The lower panel is reproduced in the main
    text. (c) Examples of states corresponding to the lowest (left),
    intermediate (middle) and highest (right) ZPEs for $a=1.5$. These
    states have uniform densities,
    $\varphi_{i\alpha}=e^{i\theta_{i\alpha}}/\sqrt{N}$.} \label{fig.sup_zpes}
\end{figure}

ZPEs for states corresponding to random three-colorings of the square
graph are shown in Fig.~\ref{fig.sup_zpes}a for different values of
$a$. The total range of ZPEs, ${\rm max}[\Lambda]-{\rm min}[\Lambda]$,
decreases rapidly when the limit $a\to 2$ is approached (see
Fig.~\ref{fig.sup_zpes}b). This quantity gives an estimate of the
magnitude of thermal fluctuations that would destabilize the
condensate~\cite{You2012}, and this is thus a signal that the
condensate becomes increasingly unstable close to $a=2$. As mentioned
in the main text, this is corroborated by the condensate fraction
decreasing close to this limit. The instability at $a=2$ is explained
in more detail in Sec.~\ref{sec.a2}. 

Some examples of $\ket{\varphi_0}$ corresponding to the lowest,
intermediate and highest ZPEs are shown in Fig.~\ref{fig.sup_zpes}c.
For all values of $a$, we find that the states $\ket{\varphi_0}$ with
the lowest zero-point energy correspond to two-colorings of the
square lattice. This means that $\omega_i$ alternates between two
values, e.g. $1/\sqrt{N}$ and $e^{i\theta}/\sqrt{N}$ with
$\theta=2\arcsin(a/2)$.

This state corresponds to a superposition of two Bloch
states. The Bloch states at $\vec{k}=(0,0)$ and
$\vec{k}=(\pi,\pi)$ have periodic parts $\ket{u_{\Gamma}}=(1,0,0)^T$
and $\ket{u_{M}}=(a/2,-1,-1)^T/\sqrt{2+(a^2/4)}$, respectively. From
these states, it is possible to construct a uniform-density state
$(\sqrt{2+(a/2)^2}\ket{u_{M}}e^{i(\pi,\pi)\cdot\vec{R}_i}\pm i
\sqrt{4-a^2}\ket{u_{\Gamma}}/2)/\sqrt{3}$, where $\vec{R}_i$ is the
position of the $i$:th unit cell. This is the superposition
corresponding to two-colorings of the square lattice.

The highest zero-point energy is found for three-colorings where the
color index always increases or decreases when moving along $x$,
$y$. These states correspond to the Bloch functions at $\vec{k}=2{\rm
  arcsin}(a/2)(\sigma,\sigma')$, $\sigma,\sigma'\in\{-1,1\}$, which
have uniform densities. In this model, the most favorable states for
condensation are therefore superpositions of Bloch functions, while
Bloch functions with uniform densities are the least favorable. A
standard momentum-space approach would therefore consider condensation
in the least favorable uniform-density state, except if the unit cell
is enlarged. 

\subsubsection{Instability of condensation for $a=2$} \label{sec.a2}

\begin{figure}
  \includegraphics[width=\columnwidth]{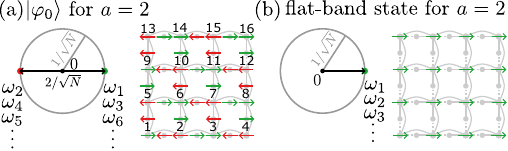}
  \caption{(a) Bloch state at $(\pi,\pi)$ and corresponding
    coefficients $\omega_i$ in the complex plane. When $a=2$, this is
    the only flat-band eigenstate minimizing $E_{\rm MF}$. (b)
    Flat-band eigenstate which contributes to the destabilization of
    the condensate for $a=2$. This state does not minimize $E_{\rm
      MF}$, but its existence still contributes to destabilizing the
    condensate due to geometric effects. }\label{fig.sup_lieb}
\end{figure}

As shown in the main text, bosonic condensation becomes unstable in
the Tasaki lattice when approaching $a=2$. At $a=2$, the distance
between coefficients $\omega_i$ corresponding to overlapping CLSs
needs to be $2/\sqrt{N}$ for uniform densities. Since the coefficients
are all on the same circle of radius $1/\sqrt{N}$, they need to be
diametrically opposed: there is only one $\ket{\varphi_0}$ minimizing
the mean-field energy (excluding gauge transformations), shown in
Fig.~\ref{fig.sup_lieb}(a). This is the Bloch state at
$(\pi,\pi)$. Because there is no degeneracy in the minimization of
$E_{\rm MF}$, we could naively expect the condensate to be more stable
than at $a<2$, not less.

However, the issue here is the existence of a multitude of non-uniform
flat-band eigenstates relating to $\ket{\varphi_0}$ by
$\ket{\varphi_0'}=R\ket{\varphi_0}$ (one example is shown in
Fig.~\ref{fig.sup_lieb}(b)). In the complex plane, if we do not
require uniform densities, we can move every coefficient $\omega_i$
continuously along the diameter of the circle without adding complex
phases to any $\varphi_{i\alpha}$. While the
framework consists of only triangles with $0$ and two coefficients
$\omega_i$ as their vertices, the area of the triangles vanishes, and
stability is no longer guaranteed.

\end{document}